\newif\ifpdflatex    
\pdflatextrue           

\documentclass[fleqn,usenatbib,useAMS]{mnras}

\usepackage{graphicx}	
\usepackage{amsmath}	
\usepackage{amssymb}	
\usepackage{multicol}        
\usepackage{bm}		
\usepackage{pdflscape}	
\usepackage{url}
\usepackage{amsbsy}
\usepackage{xspace}
\usepackage{hyperref}
\usepackage{longtable}
\usepackage[colorinlistoftodos]{todonotes}
\usepackage[flushleft]{threeparttable}
\usepackage[T1]{fontenc}
\usepackage{ae,aecompl}

\hypersetup{linkbordercolor=green}

\def\objname{M31-LRN-2015\xspace}

\def\lesssim{\mathrel{\hbox{\rlap{\hbox{\lower5pt\hbox{$\sim$}}}\hbox{$<$}}}}
\def\gtrsim{\mathrel{\hbox{\rlap{\hbox{\lower5pt\hbox{$\sim$}}}\hbox{$>$}}}}
\def\kms{{\rm km\,s$^{-1}$}\xspace}

\def\Lsun{{\rm L$_{\odot}$}\xspace}
\def\Rsun{{\rm R$_{\odot}$}\xspace}
\def\Msun{{\rm M$_{\odot}$}\xspace}

\def\halpha{{\rm H$\alpha$}\xspace}
\def\hbeta{{\rm H$\beta$}\xspace}

\newcommand{\angstrom}{\textup{\AA}\xspace}
\newcommand{\io}[2]{#1\,{\textsc{#2}}}

\def\aj{AJ}%
\def\apj{ApJ}%
\def\apjl{ApJ}%
\def\apjs{ApJS}%
\def\aap{A\&A}%
\def\aapr{A\&A~Rev.}%
\def\mnras{MNRAS}%
\def\pasp{PASP}%
\def\nat{Nature}%
\def\procspie{Proc.~SPIE}%
\def\pasa{PASA}%
\def\apss{Ap\&SS}%

\title[M31 Stellar Merger]{Progenitor, Precursor and Evolution of the Dusty Remnant of the Stellar Merger \objname}

\author[Blagorodnova et al.]{
\parbox{17.5cm}{
  N.~Blagorodnova,$^{1}$ \thanks{E-mail: n.blagorodnova@astro.ru.nl} 
  V.~Karambelkar,$^{2}$
  S.~M.~Adams,$^{2}$
  M.~M.~Kasliwal,$^{2}$
  C.~S.~Kochanek,$^{3,4}$
  S.~Dong,$^{5}$
  H.~Campbell,$^{6}$
  S.~Hodgkin,$^{7}$
  J.~E. Jencson,$^{8}$
  J.~Johansson,$^{9}$
  S.~Koz{\l}owski,$^{10}$
  R.~R.~Laher,$^{11}$
  F. Masci,$^{11}$
  P.~Nugent,$^{12}$\\
U.~Rebbapragada $^{13}$
  }
  \\
  \\
  $^{1}$ Department of Astrophysics/IMAPP, Radboud University, Nijmegen, The Netherlands\\
  $^{2}$ Cahill Center for Astrophysics, California Institute of Technology, Pasadena, CA 91125, USA\\
  $^{3}$ Department of Astronomy, The Ohio State University, 140 W. 18th Ave., Columbus, OH 43210, USA\\
  $^{4}$ Center for Cosmology and AstroParticle Physics (CCAPP), The Ohio State University, 191 W. Woodruff Ave., Columbus, OH 43210, USA\\
   $^{5}$ Kavli Institute for Astronomy and Astrophysics, Peking University, Yi He Yuan Road 5, Hai Dan District, Beijing 100871, China\\
    $^{6}$ Department of Physics, Faculty of Engineering and Physical Sciences, University of Surrey, Guildford, Surrey, GU2 7XH, UK \\
   $^{7}$ Institute of Astronomy, University of Cambridge, Madingley Road, CB3 0HA, Cambridge, UK \\
   $^{8}$ Steward Observatory, University of Arizona, 933 North Cherry Avenue, Tucson, AZ 85721-0065, USA\\
  $^{9}$ Department of Physics and Astronomy, Division of Astronomy and Space Physics, Uppsala University, Box 516, SE-751 20 Uppsala, Sweden\\
  $^{10}$ Astronomical Observatory, University of Warsaw, Al. Ujazdowskie 4, 00-478 Warszawa, Poland\\
  $^{11}$ Infrared Processing and Analysis Center, California Institute of Technology, Pasadena, CA 91125, USA\\
  $^{12}$ Lawrence Berkeley National Laboratory, Berkeley, CA 94720, USA \\
  $^{13}$ Astrophysics Research Institute, Liverpool John Moores University, 146 Brownlow Hill, Liverpool, L3 5RF, UK\\
  $^{14}$ Jet Propulsion Laboratory, California Institute of Technology, Pasadena, CA 91125, USA \\
    }

\date{Last updated \today; in original form \today}

\pubyear{2020}

\begin{document}
\label{firstpage}
\pagerange{\pageref{firstpage}--\pageref{lastpage}}
\maketitle

\begin{abstract}
\objname is a likely stellar merger discovered in the Andromeda Galaxy in 2015. We present new optical to mid-infrared photometry and optical spectroscopy for this event. Archival data shows that the source started to brighten $\sim$2 years before the nova event. During this precursor phase, the source brightened by $\sim$3\,mag. The lightcurve at 6 and 1.5\,months before the main outburst may show periodicity, with periods of 16$\pm$0.3 and 28.1$\pm$1.4\,days respectively. This complex emission may be explained by runaway mass loss from the system after the binary undergoes Roche-lobe overflow, leading the system to coalesce in tens of orbital periods. While the progenitor spectral energy distribution shows no evidence of pre-existing warm dust in system, the remnant forms an optically thick dust shell at $\sim$4 months after the outburst peak. The optical depth of the shell increases dramatically after 1.5\,years, suggesting the existence of shocks that enhance the dust formation process. We propose that the merger remnant is likely an inflated giant obscured by a cooling shell of gas with mass $\sim0.2$\,\Msun ejected at the onset of the common envelope phase. 
\end{abstract}

\begin{keywords}
 binaries (including multiple): close -- stars: evolution -- (stars:) novae, cataclysmic variables -- (ISM:) dust, extinction -- ISM: jets and outflows
\end{keywords}

\section{Introduction}
Luminous red novae (LRNe) are an observational class of astrophysical transients with peak luminosities in between novae and supernovae ($-3<M_{V}<-14$\,mag), making them members of the ``gap transients'' population \citep{Kasliwal2012PASA,Pastorello2019NatAs}. This population includes transients of very diverse nature, such as faint core-collapse supernovae (SNe), .Ia-like SN explosions \citep{Bildsten2007ApJ}, Ca-rich (strong) transients \citep{Kasliwal2012ApJ}, LRNe \citep{Kulkarni07}, 2008S-like objects  \citep{Prieto09,Thompson09,Botticella2009MNRAS} and LBV-like outbursts \citep{Smith2011MNRAS} among others. Within the gap category, there is an observational class characterized by strong interaction with a circumstellar medium (CSM), called Intermediate Luminosity Optical/Red Transients (ILOT/ILRT). LRNe belong to this group. Observationally, LRN transients quickly evolve towards colder (redder) temperatures, resembling K or M-type giant stars. Their spectra are characterized by hydrogen emission lines expanding at velocities $\lesssim1000~\mathrm{km}\>\mathrm{s}^{-1}$ and a forest of absorption lines for low ionization elements. At later stages, the progressively cooling photosphere shows the formation of molecules and new dust in the system, which quickly obscures the optical emission \citep{Kaminski2011AA,Kaminski2015AA}.

Since the discovery of the first ``red variable'' about 30\,years ago \citep[M31RV;][]{Rich89,Mould1990ApJ}, LRNe have been sporadically detected both in the Milky Way and in galaxies within $\sim30$\,Mpc \citep{Kulkarni07,Smith16,Mauerhan2018MNRAS,Blagorodnova17,Pastorello2019a}. Although their fainter luminosities pose a challenge to their discovery, a few nearby well studied cases provided a great wealth of information about their origin as violently interacting binary systems. The best example to date is V1309\,Sco \citep{Mason2010AA}. Archival data revealed it to be a contact eclipsing binary, whose period decreased exponentially for several years prior to the nova outburst \citep{Tylenda11}. Other members of this class include the Galactic stellar mergers V838\,Mon \citep{Munari2002AA}, V4332\,Sgr \citep{Martini99} and OGLE-2002-BLG-360 \citep{Tylenda13}.

LRNe provide unique observational clues about the final catastrophic stages of binary interaction, specially the termination of the so-called common envelope (CE) phase. This phase can be initiated by the Darwin instability \citep{Darwin1879RSPS}. The spin up of the primary component in an initially tidally-locked binary would extract angular momentum from the orbit, drive the orbital decay, and ultimately initiate Roche lobe overflow \citep{Rasio1995ApJ}. The donor star (typically the primary evolving off of the main-sequence) overfills its Roche lobe and starts unstable mass transfer towards its companion \citep[see][for a detailed explanation]{Izzard2012IAUS}. This process may culminate with both stars orbiting inside a shared non co-rotating layer of gas, called the common envelope \citep{Paczynski1976IAUS}. The less massive component quickly spirals inwards, transferring the angular momentum of the binary to the envelope. At the termination of this phase part (or all) of the envelope can be ejected, leaving a more compact binary, or a fully coalesced star. 

The light curves of LRNe are potentially powered by three different mechanisms: free expansion and release of thermal energy from hot gas created during the dynamical phase of the merger (similar to ``cooling envelope'' emission in supernovae); recombination of ionized ejecta \citep{Ivanova13}, analogous to SNe IIP, and sustained luminosity due to reprocessed shock emission as the dynamical ejecta runs into surrounding layers of pre-existing gas \citep{Metzger2017}. 

Well-studied stellar mergers are powerful probes of the dynamical onset of common envelope evolution, which is one of the most challenging phases of binary evolution \citep[see][for a review]{Ivanova2013araa}. Currently, there are still important questions to be answered. Which progenitor systems undergoing Roche lobe overflow (RLOF) result in CE events via dynamically-unstable mass transfer? What is the role of pre-outburst mass loss from the system \citep{Pejcha14,Pejcha16b,Pejcha16,MacLeodOstriker2018ApJ,MacLeodOstrikerStone2018ApJ,Reichardt2019MNRAS,MacLeodLoeb2020ApJ}? What is the role of jets in the removal of the envelope \citep{MorenoMendez2017MNRAS,ShiberSoker2018MNRAS,Shiber2019MNRAS,Lopez-Camara2020arXiv}?
How dust forms in the expanding ejecta \citep{Lu2013ApJ,Iaconi2020arXiv} and what is the role of dust-driven winds in unbinding the loosely bound envelope after its ejection \citep{Glanz18}? Finally, understanding which common envelope events result in a complete merger and which result in a close binary binary is crucial for gravitational wave science \citep{Dominik2012ApJ,VignaGomez2020arXiv,Klencki2020arXiv200611286K}. 

The rate of transients associated to stellar mergers is high enough \citep{Kochanek14_mergers} to ensure several detections per year with ongoing surveys, such as the Zwicky Transient Facility  \citep[ZTF;][]{Adams2018,Howitt2020MNRAS}. However, few are likely to happen as close as M31. The event we discuss here, \objname, is probably the closest extragalactic example of this transient family for the next 20 years, and thus provides the best opportunity for studying the evolution of the remnant at late times.

\objname was discovered in the Andromeda galaxy (M31) at $\alpha_{\rm{J2000}}=00^{\rm{h}}42^{\rm{m}}07^{\rm{s}}.99$, $\delta_{\rm{J2000}}=+40^{\circ}55^{'}01^{''}.1$, (6.8$'$ W and $-21.13'$ S of its galaxy centre) on UTC 2015 January 13.63 by the MASTER survey \citep{Shumkov15}.  Though the transient was initially classified as a classical nova peaking at $R\sim15.1$ mag, it quickly evolved towards red colours and was re-classified as a LRN based on its luminosity, slow decline, color evolution, and spectroscopic similarities to V838 Mon by \citet{ATel6941}.  \citet{Dong15} identified a progenitor in archival Hubble Space Telescope (\emph{HST}) imaging and noted that the progenitor had started brightening in later archival Canada France Hawaii Telescope (CFHT) photometry.  \citet{Kurtenkov15}, \citet{Williams2015}, and \citet{Lipunov17} presented optical and near infrared (NIR) light curves and optical spectra of the outburst spanning from $\sim$10\,days pre-maximum to $\sim$60\,days post-maximum.

Further analysis by \citet{MacLeod17} used the archival photometry from \citet{Dong15} and \citet{Williams2015} to estimate the evolutionary stage of the progenitor star for single stellar models. The best agreement was for a $3-5.5$\,\Msun sub-giant star with a radius $30-40$\,\Rsun that was evolving off the main sequence. In their model, the optical transient lasted less than 10 orbits of the original binary, and the light curve was best explained by $\sim 10^{-2}$\,\Msun of fast ejecta driven by shocks at the onset of CE followed by H recombination of an additional $\sim 0.3$\,\Msun of material ejected at slower velocities as the secondary continued to spiral in through the envelope of the primary.

Similar results were obtained by \citet{Lipunov17}, using hydrodynamic simulations designed to reproduce the multi-colour light curve of the transient. The observed $\sim$50\,day plateau was explained by a stellar merger with a progenitor mass of 3\,\Msun and a radius of 10\,\Rsun, which underwent fast shock heating at the base of the envelope accelerating the matter to photospheric velocities of $v \sim 900$\,\kms. In this scenario, the system is already a tight binary with a degenerate dwarf primary, and the unstable mass transfer is initiated by the secondary companion.

Alternatively, \citet{Metzger2017} suggested that the \objname had a more extended period of mass loss from the L2 point before the outburst, which powered the emission observed prior to the main peak. The light curve could then be explained by the release of thermal energy near the peak, followed by a plateau powered by the interaction of a dynamically ejected shell with the preexisting equatorial material. 

In this paper we present new optical and infrared photometry, and optical spectroscopic data on \objname{}. We investigate the evolution of the progenitor up to 5.5\,years before the outburst, which shows a more complex behaviour. We also provide the first analysis of the remnant, based on infrared photometry taken up to 5\,years after the outburst. We present new observational data in Section \ref{sec:observations}. Our analysis of the object's spectroscopic and photometric evolution is in Section \ref{sec:analysis}. We discuss the implications of our results in Section \ref{sec:discussion} and summarize our conclusions in Section \ref{sec:conclusions}.

\section{Observations} \label{sec:observations}

\subsection{Distance and reddening}

Following \citet{Williams2015} and \citet{Lipunov17} we adopt a distance of 0.762\,Mpc, corresponding to a distance modulus of $(m-M)=24.4$\,mag \citep{Freedman1990}.

There is some uncertainty in the reddening that should be adopted for \objname.  Based on the foreground reddening of $E(B-V)=0.062$\,mag \citep{Schlegel98} and the total line-of-sight extinction at the position of \objname of $E(B-V)=0.18$ \citep{Montalto2009}, \citet{Williams2015} and \citet{Lipunov17} assume \objname was reddened by $E(B-V)=0.12\pm0.06$\,mag.  \citet{Kurtenkov15} report a line of sight reddening of $E(B-V)=0.42\pm0.03$\,mag based on dust maps of M31 \citep{Draine2014}, and find $E(B-V)=0.35\pm0.10$ mag from modeling spectra taken between UTC 2015 January 15 and 2015 February 24.  In this paper we adopt a foreground extinction of $E(B-V)=0.055$ based on the \citet{Schlafly12} recalibration of the \citet{Schlegel98} dust maps plus an average host extinction of $E(B-V)=0.2$, which is a compromise between all previous studies. Therefore, the total extinction used in this study is $E(B-V)=0.255$.  

\subsection{Photometry} \label{sec:photometry}

We monitored the photometric and spectroscopic evolution of \objname with a variety of ground-based facilities and with the \textit{Spitzer Space Telescope}  \citep[SST;][]{Werner2004,Gehrz2007}. We also collected all the published data on this source from literature. A full log of archival and follow-up photometry is shown in Table \ref{tab:photometry}.

\begin{table}
\begin{center}
\begin{minipage}{0.5\textwidth}
\caption{Archival (see references) and follow-up photometry for \objname. These values have not been corrected for extinction. 
This table is shown as a guidance on the format. The full table is available as part of the online material.}
\label{tab:photometry}
\begin{tabular}{rllrrr}
\hline
\hline
{Phase} & {MJD} & {Telescope} & {Filter} & {Mag} & {ATel} \\ 
\hline
-3810.7 & 53233 & HST        & I      & 22.02$\pm$0.01  & 7173        \\
-3810.7 & 53233 & HST        & V      & 23.2$\pm$0.03   & 7173        \\
-3443.7 & 53600 & Spitzer      & 3.6      &       $>$17.99        &         \\
-3443.7 & 53600 & Spitzer      & 4.5      &       $>$17.83        &         \\
-3443.7 & 53600 & Spitzer      & 5.8      &       $>$17.03        &         \\
-3443.7 & 53600 & Spitzer      & 8.0      &       $>$15.89        &         \\
... & ... & ... & ... & ... & ... \\
1260.8    & 58304.5 & Keck       & K      & $>$24.26  &                 \\
1291.8    & 58335.5 & Keck       & J      & $>$25.27  &                 \\
1291.8    & 58335.5 & Keck       & H      & $>$24.81   &  \\
1793.2    & 58836.9 &  Spitzer  & 3.6   &  $>$17.2 & \\
1793.2    & 58836.9 &  Spitzer  & 4.5 & 15.73$\pm$0.02 & \\
\hline
\hline
\end{tabular}
\begin{tablenotes} \item References: ATel 7173 \citet{Dong15} \end{tablenotes}
\end{minipage}
\end{center}
\end{table}

The precursor and the nova photometry were retrieved from archival images from the Palomar Transient Facility \citep[PTF;][]{Law09,Rau09} and the Intermediate Palomar Transient Facility \citep[iPTF;][]{Kulkarni13}. The images were obtained between MJDs 55569.0 and 57642.0 (5.3\, prior to discovery to 1.6 years after) with the CFH12K camera \citep{Rahmer08,Law10} on the Palomar 48-inch Telescope (P48).  The PTF/iPTF data were processed with the PTF image subtraction pipeline \texttt{PTFIDE} \citep{Masci17}. 
The CFHT/MegaPrime light curves with $\sim$6 and $\sim$26\,min cadence were obtained from three full nights monitoring the field for microlensing on 2014 October 24, 28 and 30 (PI: Subo Dong). The photometry was obtained using difference imaging analysis \citep{Wozniak2000AcA} and calibrated using the Pan-STARRS DR1 \citep[PS1;][]{Chambers16}. 

Optical follow-up imaging was obtained with the Liverpool Telescope \citep{Bersier15} and with the Large Binocular Camera \citep[LBC;][]{Giallongo08} on the Large Binocular Telescope \citep[LBT;][]{Hill06}.

\begin{figure}
\includegraphics[width=0.47\textwidth]{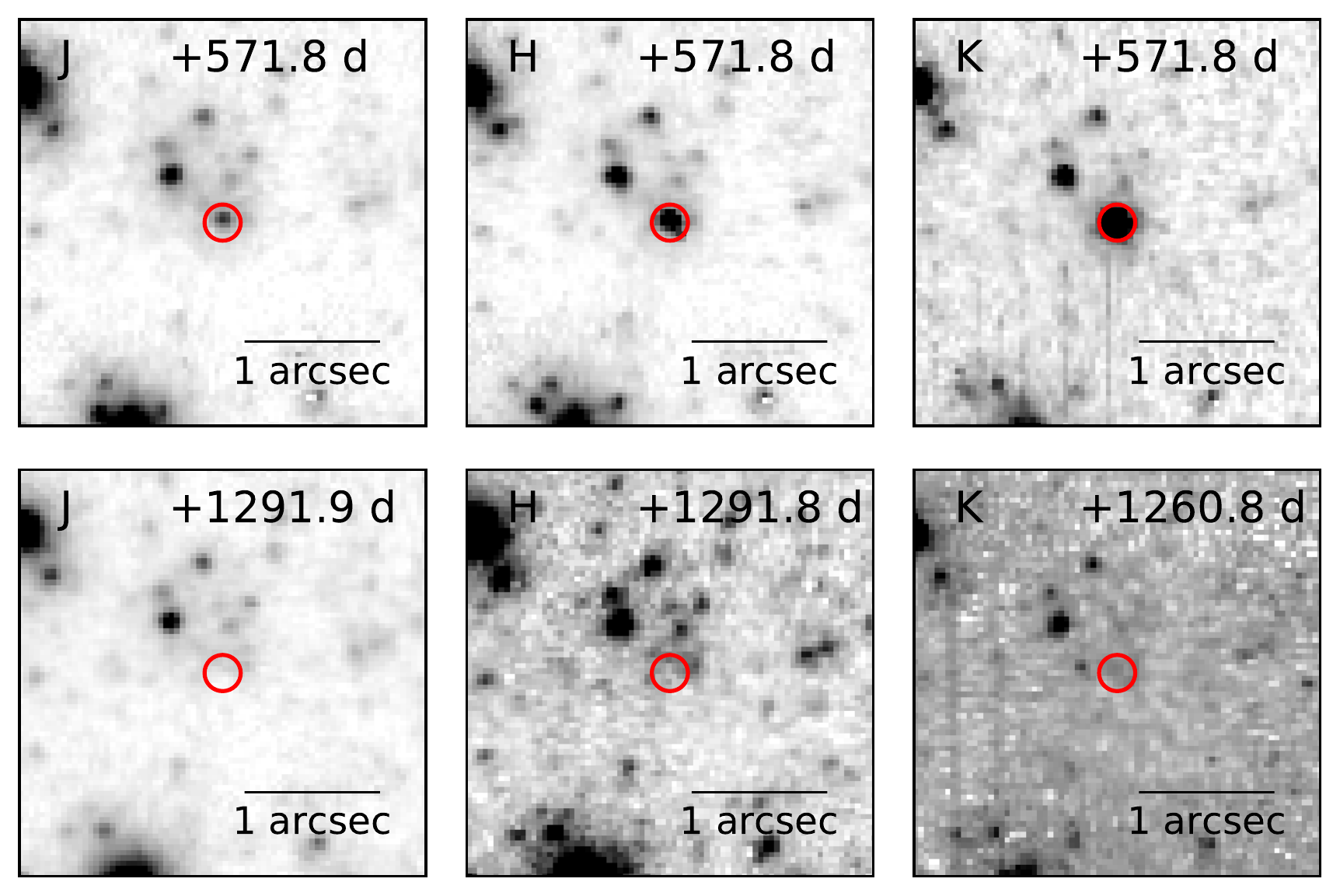}
\caption{Late time follow-up of \objname in $J, H, Ks$ bands with Keck II/NIRC2-AO. The location of the target is indicated with a red circle. The remnant is clearly detected at 571.8\,days ($\sim$1.5\,years), but at 1291.9\,days the star has faded beyond the confusion limit. The cutouts have the standard orientation North:up, East:left.} \label{fig:nir_followup}
\end{figure}

Near-IR imaging was collected with the LBT NIR-Spectroscopic Utility with Camera and Integral-Field Unit for Extragalactic Research \citep[LUCIFER;][]{Seifert03}, with the CAmara INfrarroja (CAIN) on the Telescopio Carlos S{\'a}nchez and, at late times, with the NIRC2-AO camera on the 10\,m Keck II telescope. The observations were made with $J, H$ and $K_s$ filters in the wide camera mode, having 40\,mas pixel$^{-1}$. We used the laser guide star adaptative optics \citep[LGS AO;][]{Wizinowich2006PASP} system to obtain diffraction limited images with an FWHM of approximately 2.5 pixels (equivalent to 0.1 arcsec), required to resolve the fading progenitor in a crowded M31 star field. Figure \ref{fig:nir_followup} shows the location of the outburst 1.5 and 3.5\,years after discovery. 

\begin{figure*}
\includegraphics[width=\textwidth]{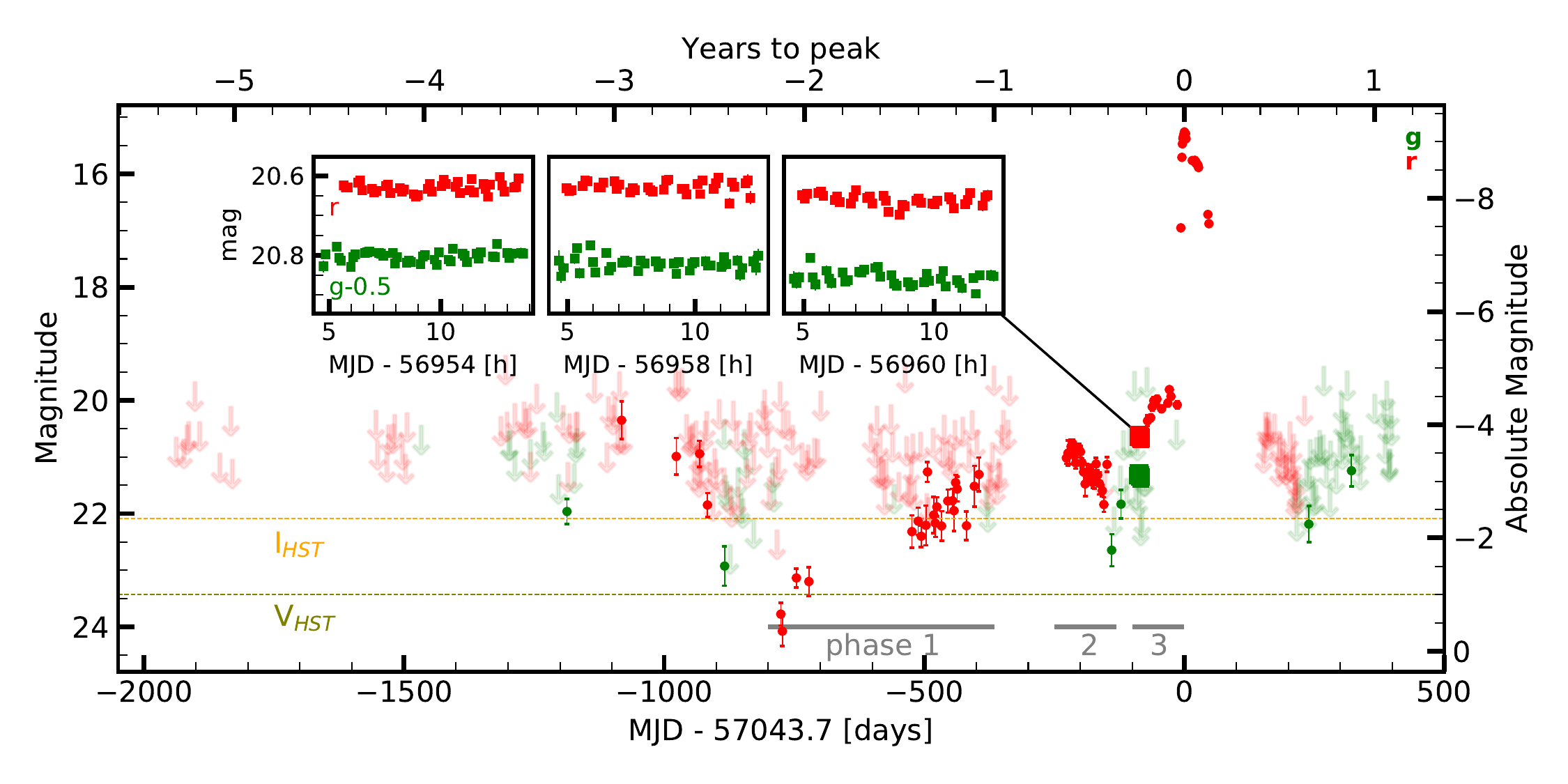}
\caption{PTF/iPTF light curve of \objname in Mould-$R$ (red) and $g$ bands (green). The light curve earlier than $-$20\,days has been binned using a bin size of 3 days. At later dates, bins of 1 days were used. The HST progenitor limits at MJD 53233 ($-$10.4\,years) reported by \citet{Dong15,Williams2015} are shown as dashed horizontal lines. The green and red squares show the CFHT photometry.  The measurement of the progenitor is reported in \citet{Dong15}. The insets show high cadence CFHT observations of the field on three full nights at $-$90, $-$86 and $-$64\,days before peak. Gray bars in the bottom indicate the different phases of the precursor lightcurve. \label{fig:ptflc}} 
\end{figure*}

Confusion complicates the interpretation of ground-based (seeing-limited) photometry of the progenitor and remnant. Archival images from Hubble Space Telescope (\emph{HST}) obtained on UTC 2004 August 16 also reveal multiple sources within 1$\arcsec$ of the progenitor with comparable fluxes \citep{Dong15,Williams2015}.  Image subtraction via the \texttt{PTFIDE} pipeline for the PTF data and via \texttt{ISIS} for the LBT data enables accurate measurement of changes in flux, but not the total flux of the target. However, due to the faintness of the progenitor, we assume that the deviation from the total flux is negligible. The difference imaging precursor light curve is shown in Figure \ref{fig:ptflc}. 

\begin{figure*}
\includegraphics[width=\textwidth]{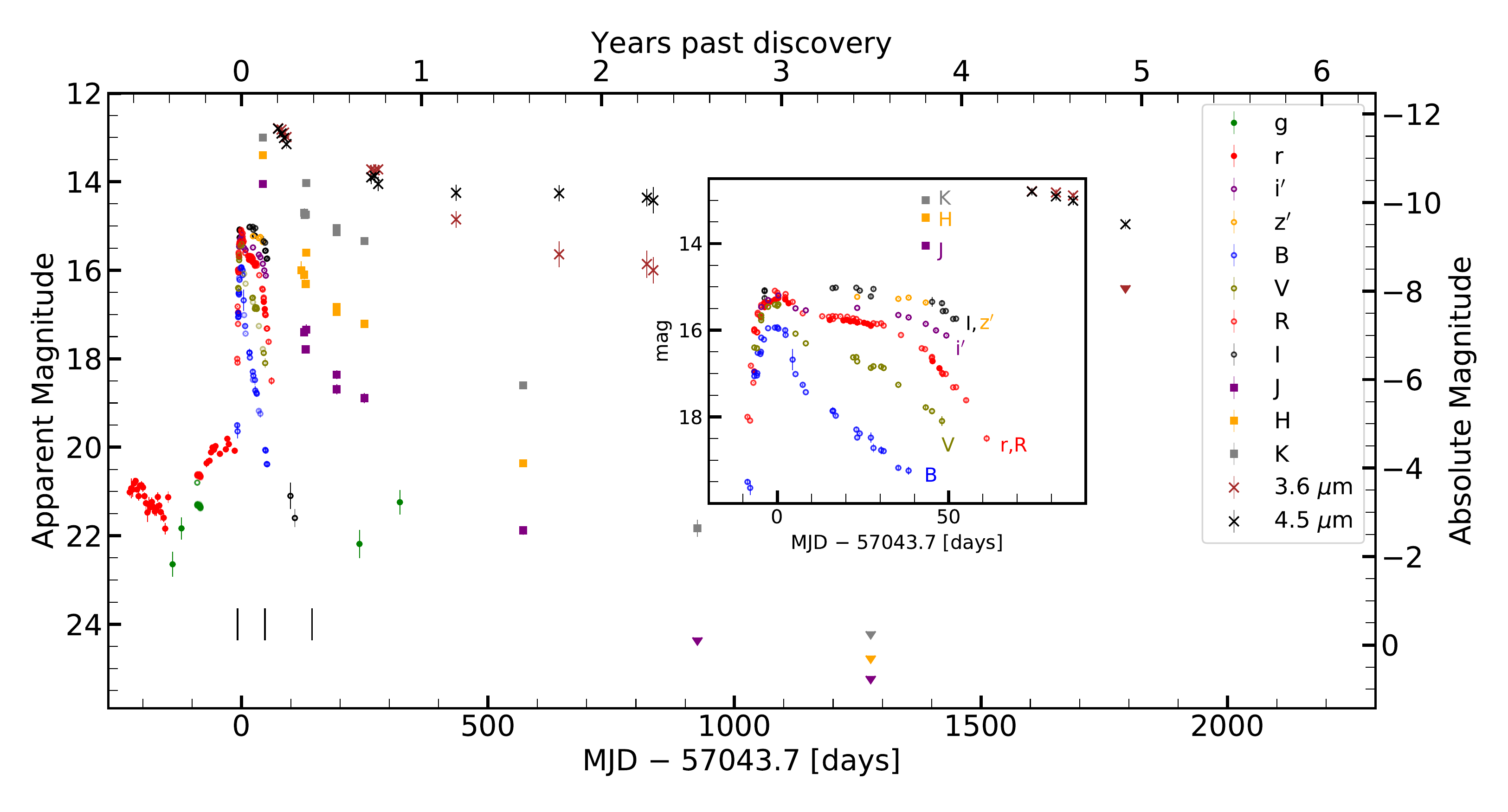}
\caption{Light curve showing PTF/iPTF photometry, along with our NIR and MIR data. For better clarity, we added the measurements published by \citet{Williams2015} and \citet{Kurtenkov15} and ATels \citep{ATel6924, ATel6941, ATel6951, ATel7236, ATel7272, ATel7468, ATel7485, ATel7555, ATel7572, ATel7595, ATel7624, ATel8059, ATel8220}. Downward triangles show upper limits. The inset shows the light curve around peak. The vertical marks under the light curve show the times when our follow-up spectra were obtained. The complete photometry is given in Table  \ref{tab:photometry}.} \label{fig:m31lc}
\end{figure*}

In our mid-IR follow-up (Figure \ref{fig:m31lc}), we utilize both new and archival mid-IR data in [3.6] and [4.5] bands from the Infrared Array Camera \citep[IRAC;][]{Fazio2004} on board  \emph{Spitzer Space Telescope}. Under programs 11181, 12063 PI: C. Kochanek), and 13053, 14089 (PI: M. Kasliwal), we obtained mid-IR imaging from UTC 2015 April 06 to UTC 2019 December 19). We also analyzed the archival images from UTC 2005 August 18/19 (program 3126; PI: P. Barmby)\footnote{We do not use archival imaging from this program taken on 2005-01-20/21 because a solar proton event flooded the detector with cosmic rays.}, and from 2015 taken under the program 11103 (PI: O. Jones). The \textit{Spitzer} photometry was reduced using the difference imaging pipeline developed for the \textit{Spitzer} InfraRed Intensive Transients Survey (SPIRITS) survey \citep{Kasliwal2017ApJ}.

\subsection{Spectroscopy} \label{sec:spectroscopy}

We acquired optical follow-up spectra with the OSMOS spectrograph on the MDM Observatory 2.4\,m Hiltner telescope on UTC 2015 March 10 \citep[+47.4\,days;][]{Wagner15} and with the Low Resolution Imaging Spectrograph \citep[LRIS;][]{Oke95} on the 10\,m Keck I telescope on UTC 2015 June 13 (+142.9\,days). The OSMOS spectrum was reduced using standard \texttt{IRAF} routines. The LRIS spectrum was reducing using an \texttt{IDL} based pipeline \texttt{lpipe} developed by D. Perley \footnote{\url{http://www.astro.caltech.edu/~dperley/programs/lpipe.html}}. 

We also include early time (8 days pre-peak) spectroscopy from WHT/ACAM reported by \cite{Hodgkin2015ATel6952}, the Liverpool Telescope SPRAT Spectroscopy published in \cite{Williams2015} and spectroscopy from \citet{Kurtenkov15}, constituting the most complete spectroscopic dataset up to date. The ACAM spectrum was reduced using the usual \texttt{IRAF} procedures, and scaled using the photometry in $B$ and $R$ bands of the nova reported for the same night by \cite{Kurtenkov15}. The observation log for the unpublished spectra is provided in Table \ref{tab:speclog} and the spectral sequence is displayed in Figure \ref{fig:spectra}.

In order to correct the spectra for the systemic velocity of Andromeda, we adopt $v_{sys} =-304.5 \pm 6.8$\,\kms 
\citep{Chemin2009ApJ}. In addition, the transient is nearly aligned with the semi-major axis of M31 (6.8$'$ W and $21.13'$ S of M31 centre). Assuming that the progenitor is a relatively young star (100\,Myr for a 5\,\Msun) which still follows the galaxy rotation curve, we estimate that the star may have an additional correction, related to its rotation within the galaxy of $v_{rot}\sim$-200\,\kms, which should be considered as an upper limit for any additional correction to the velocity. 

\renewcommand{\tabcolsep}{0.21cm}
\begin{table*}
\begin{minipage}{0.9\linewidth}
\begin{small}
\caption{Log of spectroscopic observations of \objname.}
\begin{center}
\begin{tabular}{rccccccrrrr}
\hline
Phase$^a$ & MJD  & UTC & Telescope &     Slit & Exposure  & Airmass & Resolution & Hel. vel. \\ 
 (d) &   &	&+Instrument	& 	(arcsec) &	(s)	   &  & (km/s) & (km/s) \\ \hline
$-$7.8 & 57035.88 & 2015-01-13T21:05:06 & WHT+ACAM & 1.5 & 1200 & 1.25 & 880 & $-$25.6 \\
+47.4 & 57091.11 & 2015-03-10T02:44:47 & MDM + OSMOS & 1.2 & 600& 2.6 & 190 & $-$16.2\\
+142.9 & 57186.58 & 2015-06-13T13:58:18 & Keck I + LRIS &1.0 & 590 & 1.6 & 300 & 19.5\\
\hline
\end{tabular}
\end{center}
$^a$The phase is relative to $r$-band peak date with MJD 57043.7.
\label{tab:speclog}
\end{small}
\end{minipage}
\end{table*}

\section{Analysis} \label{sec:analysis}

\subsection{Spectroscopic Evolution}\label{sec:specevol}

\begin{figure*}
\includegraphics[width=1.\textwidth, angle=0]{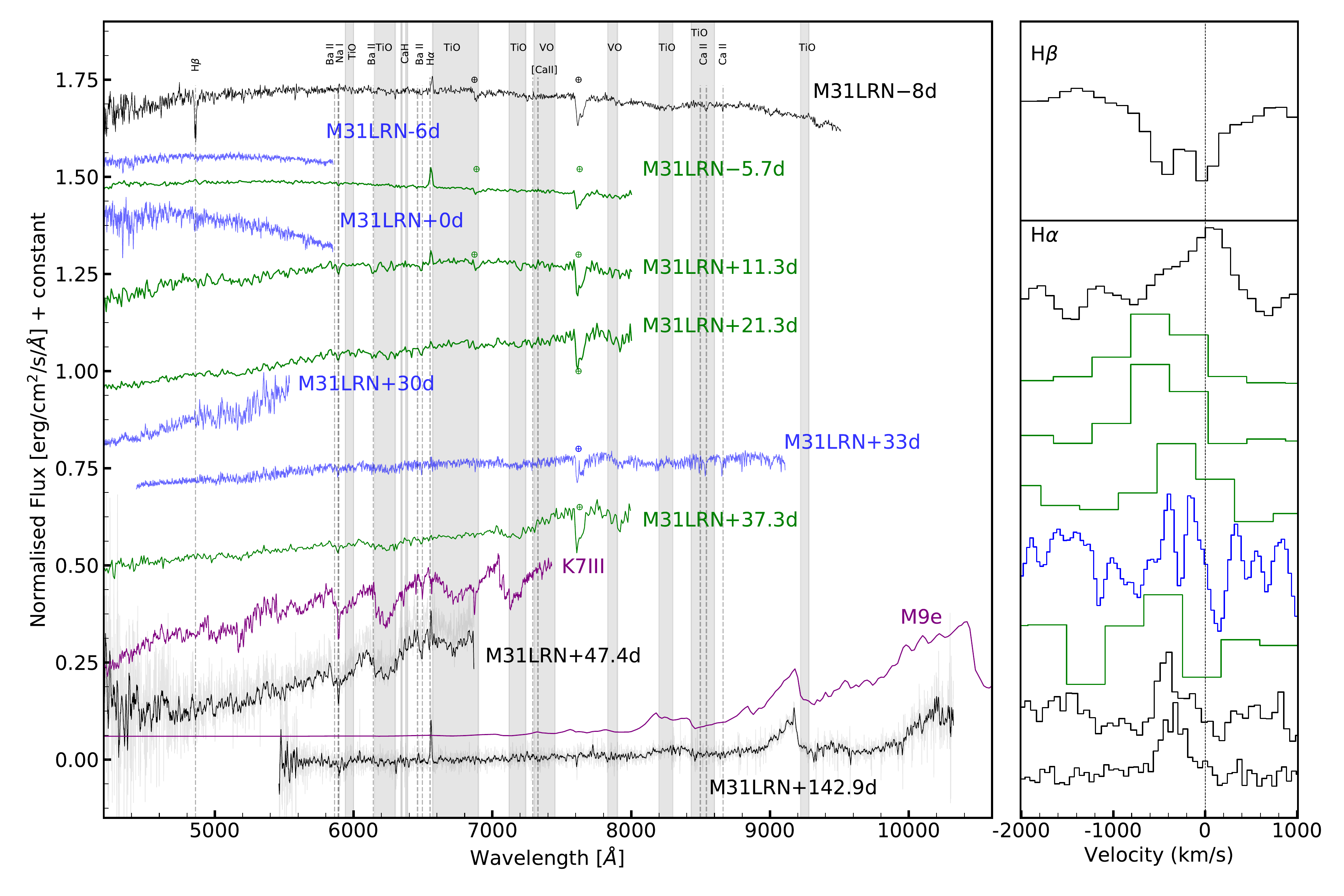}
\caption{Left: Optical spectral sequence for \objname.  The object spectra are colour coded by source. Our follow-up spectra are shown in black. A K$7$III spectrum \citep{Jacoby1984} and M9e \citep{GunnStryker1983} stellar spectra are shown in purple for comparison. As part of the sequence, we also included in green the LT spectra from \citet{Williams2015} and in blue the spectra from \citet{Kurtenkov15}. The main element lines have been identified with dashed vertical lines. Molecular absorption bands are shown with gray rectangles. Right: Continuum subtracted and peak normalized profiles for the \hbeta and \halpha lines. \label{fig:spectra}}
\end{figure*}

Our earliest spectrum was taken 8 days before optical peak. The fit for an extinction-corrected continuum shows a $\sim$5000\,K black body emission along with narrow absorption lines for low-ionization elements such as \io{Na}{i}, \io{Fe}{ii}, \io{Sc}{ii} and \io{Ba}{ii}. We also detect the \io{Ca}{ii} triplet in absorption with a velocity of $-$130\,\kms. Note the lack of emission lines for [\io{Ca}{ii}], which are consistently observed in 2008S-like transients \citep{Botticella2009MNRAS,Bond2009} (see Figure \ref{fig:spectra}). The peak of the \halpha emission is located at 6558.8\angstrom, indicating a redshift of $\sim$100\,\kms. Although the line shows a multi-component profile, the FWHM of the emission indicates that the line is likely unresolved ($<$880\,\kms). The \hbeta line is detected in absorption, centered on 4855.7\angstrom  ($-$50\,\kms) with a FWHM of 380$\pm$247\,\kms (see Table \ref{tab:balmer_lines} for the fit parameters). The profile appears to be partly filled by an emission component, in agreement with the characteristics of the spectrum at $-$6 days reported by \citet{ATel6985} and \citet{Kurtenkov15}.

The spectral sequences from \cite{Williams2015} and \citet{Kurtenkov15} show a quick progression (over $\sim$40\,days) from an almost featureless spectrum into a one dominated by strong absorption bands from recently formed molecules. The \halpha emission line detected at early times seems to disappear between +11 and +37\,days, which coincides with the duration of the light curve plateau (see Figure \ref{fig:m31lc}). If the plateau is powered by recombination, this process must be occurring deeper in the ejecta, with any emission from hydrogen recombination being absorbed by the cooler outer shell.   

Towards the end of the $r$-band plateau at +47\,days, the spectrum resembles a K7III star, with strong \io{Na}{i} and \io{Ba}{ii} absorption lines and \io{TiO}{} molecular absorption bands. The spectrum again shows the \halpha line in emission, although its peak has shifted to $\sim-300$\,\kms. This line likely comes from an expanding, cooling shell. Analogous to cool Mira stars \citep{Kaminski2017AA}, the emission is accompanied by absorption at +200\,\kms, pointing to the existence of inflowing gas.

The final spectrum, taken on the decline of the plateau, shows an almost complete lack of continuum emission blueward of 8000\angstrom, as the optical part of the spectrum has been totally absorbed by molecules, such as \io{TiO}, \io{VO}{} and \io{Zr}{O}, that are produced in O-rich atmospheres. The late time spectrum closely resembles the spectrum of the Mira star R Leo, classified at minimum as having an M9e spectral type \citep{GunnStryker1983,Keenan1974ApJS}. The \halpha line is still present in emission with a similar blueshifted velocity and signs of an absorption profile at a similar velocity.

\hspace{-0.5cm}
\renewcommand{\tabcolsep}{0.1cm}
\begin{small}
\begin{table}
\caption{Fit parameters for the \halpha and \hbeta lines. The FWHM have been corrected for instrumental profile. The line fits were corrected for heliocentric velocity and for the systemic velocity of M31.
The index ``e'' indicates the line emission component and the index ``a'' the absorption one.\label{tab:balmer_lines}}
\begin{tabular}{rcrccc} 
\hline
Phase & Ion  & Velocity$_{e}$ &FWHM$_{e}$ & Velocity$_{a}$ & FWHM$_{a}$ \\
(day)  & 	& (\kms)	& (\kms)	& (\kms)& (\kms) \\ \hline
$-$8  & \hbeta 		& -- & -- & $-48\pm 1$ & 380 $\pm$ 247 \\
$-$8  & \halpha 	& 96$\pm$30 & $<$880 & -- & --  \\
+47.4  & \halpha 	&  $-$309$\pm$1 & 115$\pm$30 & $-553\pm14$ & 136$\pm$22 \\
+142.9 & \halpha 	&  $-304\pm$1 &  300$\pm$ 20 & --  & --\\ \hline
\end{tabular}
\end{table}
\end{small}

\subsection{Spectroscopic comparison}\label{sec:discussion_speccompare}

\begin{figure*}
\includegraphics[width=1.05\textwidth]{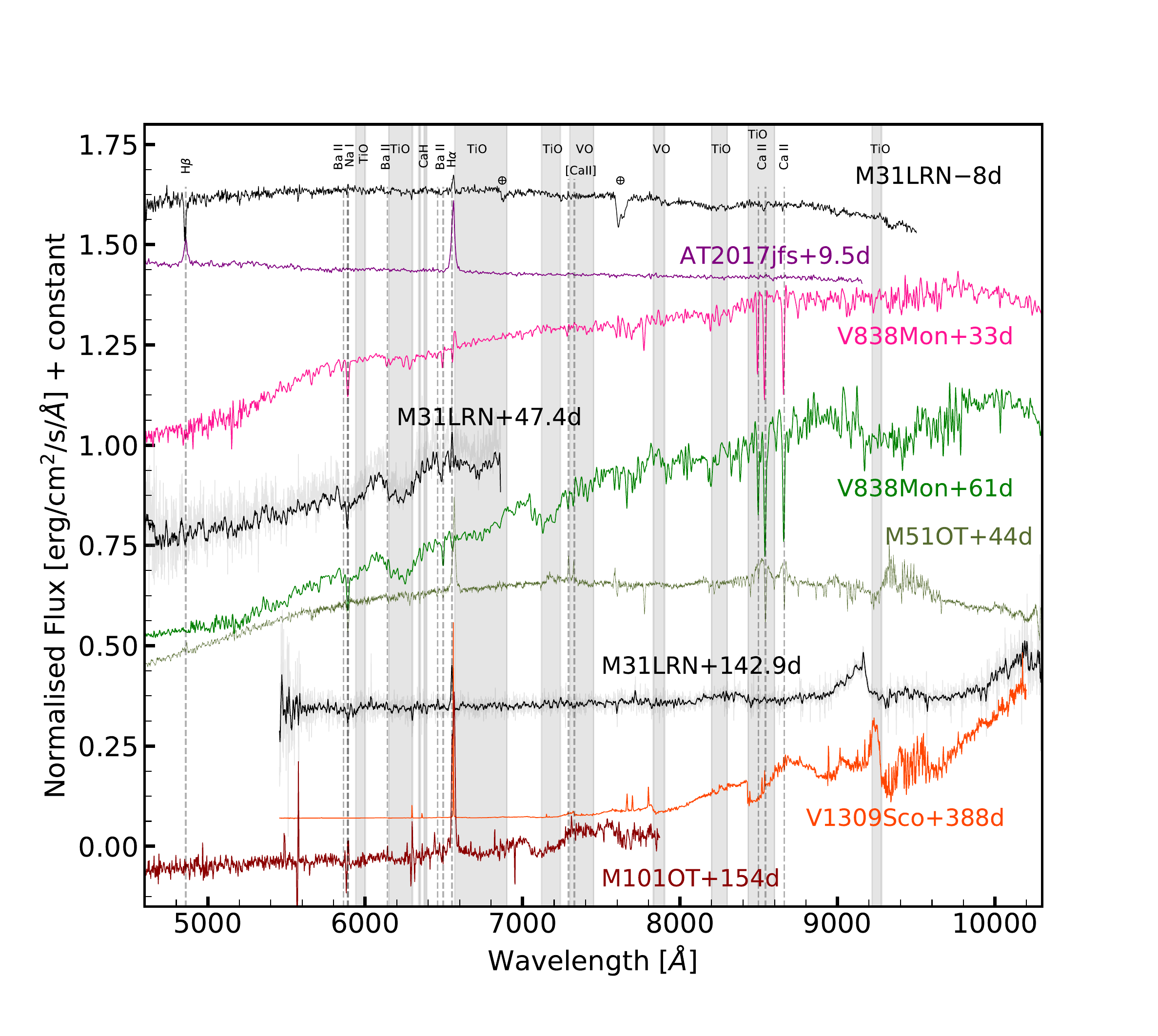}
\caption{ Spectral evolution of \objname (shown in black) compared to other ``gap'' transients at similar epochs. The spectra of the massive stellar merger AT2017jfs \citep{Pastorello2019b} during the first peak is shown in purple, the intermediate mass merger V838\,Mon \citep{Smith16} in pink and green, the 2008S-like transient M51-OT2019 \citep{Jencson2019} is shown in olive, the low-mass merger V1309\,Sco \citep{Kaminski2015AA} in orange, and the massive merger M101-OT2015 \citep{Blagorodnova17} in brown.
}
\label{fig:spectra_comparison}
\end{figure*}

In the context of stellar mergers, the spectroscopic evolution of \objname{} appears closer to the events detected in our own Galaxy than the more luminous transients recently presented by \citet{Pastorello2019a}. This last group, mainly consisting of  mergers of more massive stars, usually displayed two peaks in their light curve: a first, fast, blue peak and a second, slower red peak, which sometimes is smoothed into a pleateau. 

A comparison of the pre-peak spectra of \objname with the extragalactic transient AT2017jfs \citep{Pastorello2019b} (see Figure \ref{fig:spectra_comparison}) shows an initial difference in the continuum temperature, which is also reflected in the emission lines seen during the first peak. AT2017jfs has an initial temperature corresponding to an ionized gas ($\geq$7000\,K) and the spectrum shows signatures of interaction with an optically thick circumstellar medium (CSM), due to the strong electron scattering wings on the Balmer lines, \ion{Ca}{ii} and \ion{Fe}{ii}. On the other hand, the lower temperature of \objname{} ($\sim$5000\,K) corresponds to mostly neutral gas and shows the same elements in absorption. The \halpha and \hbeta emission is likely related to recombination in an optically thin shell, surrounding the main outburst.

The evolution towards colder temperatures is similar to V838\,Mon, where we see the appearance of \ion{TiO}{} molecular absorption features. The \halpha line, which disappeared between 11 and 37\,days, starts to become detectable again at 47\,days. We see very different signatures for spectra of 2008S-like transients taken at a similar phase. For example, M51 OT2019-1 \citep{Jencson2019} shows a much hotter continuum, and broadened Balmer emission lines due to electron scattering. These transients also show the \ion{Ca}{ii} $\lambda\lambda$ 3933, 3968 doublet in absorption, where de-excitation leads to emission in the \ion{Ca}{} NIR triplet, followed by a final forbidden transition of [\ion{Ca}{ii}] $\lambda\lambda$ 7291, 7323. This emission is not detected in LRNe.

The late time spectrum of \objname is analogous to other Galactic  \citep[V1309\,Sco;][]{Kaminski2015AA} and extragalactic \citep[M101-OT2015;][]{Blagorodnova17} stellar mergers. The cold stellar photosphere shows almost no flux at wavelengths shorter than 7000\angstrom, and the spectrum is dominated by molecular absorption features. Some low-ionization elements are detected as narrow emission lines, likely generated in an optically thin shell irradiated by unobscured emission from the central object. The increased optical depth due to formation of dust can also explain the blueshift commonly observed in the the \halpha profiles.

\subsection{Constraints on progenitor dust}\label{sec:progenitor_dust}

\begin{figure}
\includegraphics[width=0.5\textwidth]{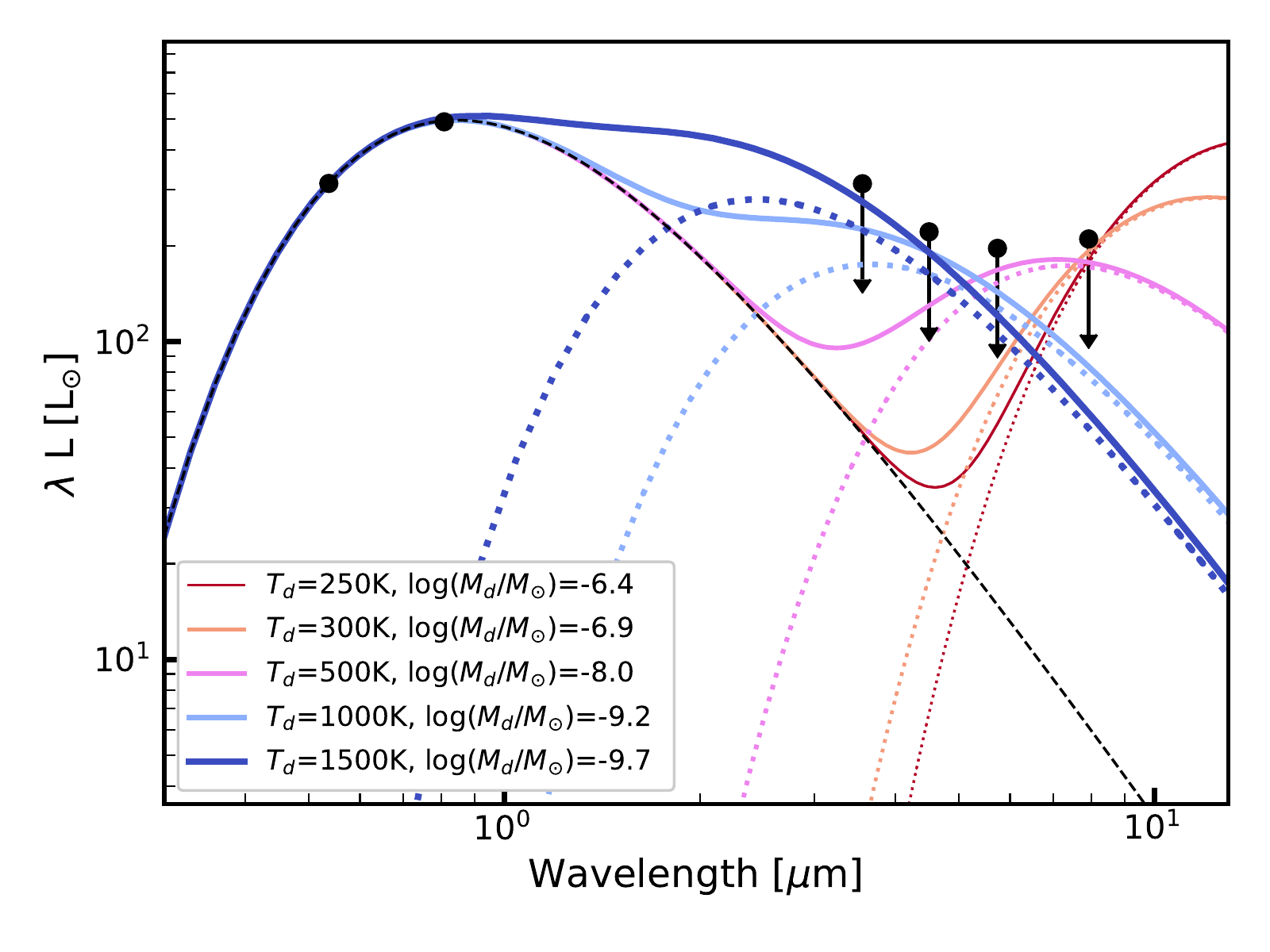}
\caption{Limits for silicate ($a=0.1\mu$m) dust emission component in the progenitor SED. The black points indicate the optical photometric measurements and the infrared upper limits. The black dashed line shows the black body fit to the progenitor photometry. The coloured dashed lines show the dust emission model at different temperatures $T_d$ for a maximum dust mass $M_d$. Solid lines of the same colour indicate the total flux emitted by the star and the dust.} \label{fig:progenitor_dust}
\end{figure}

Based on the archival SDSS and CFHT imaging data presented in \citet{Dong15}, we argue that the progenitor star had a relatively constant optical luminosity between 2002 and 2009. The photometry derived for this period agrees with the HST magnitudes $V=23.2\pm$0.03 and  $I=22.02\pm0.01$ in 2004, about 10.4\,years before the outburst peak \citep{Dong15,Williams2015}. The constant luminosity of the progenitor allows us to combine the optical data with \textit{Spitzer} upper limits obtained 9.4\,years before the outburst to place constraints on the pre-existing dust in the system.

Our simple progenitor model assumes optically thin dust of uniform grain size $a=0.1 \mu$m, which radiates at a single equilibrium temperature $T_d$. Given the abundance of oxygen-rich molecules in the ejecta of LRNe, we assume silicate dust for our analysis. The assumption of a different dust composition does not significantly alter our results. 

The specific luminosity for a dust grain of radius $a$ and temperature $T_d$ radiating at wavelength $\lambda$ is given by
\begin{equation}
l_{\nu} (\lambda) = 4 \pi a^2 B_{\nu}(\lambda, T_d) Q_{ext}(\lambda), 
\label{eq:dust_luminosity}
\end{equation}
where $B_{\nu}(\lambda, T_d)$ is the Plank function, and $Q_{ext}=Q_{abs}+Q_{sca}$ is the dust efficiency factor for extinction (absorption plus scattering). For silicate dust we use the tabulated values of \citet{Laor1993ApJ}\footnote{Available from \url{https://www.astro.princeton.edu/~draine/dust/dust.diel.html}}. Introducing the dust opacity, $\kappa(\lambda, a) = 3 Q_{ext}(\lambda) / 4 \rho_d a$, where $\rho_d$ is the dust bulk density ($\rho_d=2.2$\,g\,cm$^{-3}$ for graphitic dust and 3.5\, g\,cm$^{-3}$ for silicate), we can rewrite the equation as

\begin{equation}
l_{\nu} (\lambda) = 4 M_d \pi B_{\nu}(\lambda, T_d) \kappa(\lambda, a)   
\label{eq:dust_luminosity_kappa}
\end{equation}
where $M_d$ is the dust mass. The flux density for an observer at distance $d$ is

\begin{equation}
F_{\nu} (\lambda) = \frac{M_d B_{\nu}(\lambda, T_d) \kappa(\lambda, a)}{d^2}. \label{eq:dust_flux_density}
\end{equation}

For our analysis of the progenitor SED, we initially fit the optical HST magnitudes with a single black body to estimate its contribution to the mid-infrared flux. The best fit given our extinction is $T=4336_{-43}^{+ 41}$\,K, $R=46_{-1}^{+2}$\,\Rsun and bolometric luminosity of $L= 681_{-53}^{+64}$\,\Lsun.
Next, for each dust temperature $50 \leq T_d \leq 1500$\,K, we derive the maximum dust mass $M_d$ that is still consistent with a non-detection. 

Our results, depicted in Figure \ref{fig:progenitor_dust}, show that although the temperature has a strong influence on the dust mass, we still can place meaningful constraints for warmer dust. For example, for the hottest temperatures of $T_d=1000 - 1500$\, we find that $M_d<(10^{-9.2} - 10^{-9.7})$\,\Msun and for $T_d=500$\,K $M_d<10^{-8.0}$\,\Msun. For temperatures of 300\,K and 100\,K the peak of the emission shifts further into the far infrared, increasing the limits to $10^{-6.9}$\,\Msun and $10^{-1.7}$\,\Msun. Due to the lack of far infrared measurements, our limits for colder dust are not that constraining, so we can not rule out the presence of a cold ($\sim$30\,K) dust shell, such as the one detected for the remnant of V1309\,Sco \citep{Tylenda2016}.

\subsection{Pre-outburst progenitor evolution}\label{sec:progenitor}

The extensive coverage of M31 by CFHT and the PTF/iPTF surveys since 2010 provides enough data to study the evolution of the progenitor up to five years before its outburst. Using the \texttt{PTFIDE} pipeline, we obtained hundreds of forced photometry measurements for observations taken between 2010 and the onset of the nova in January 2015, as shown in Figure \ref{fig:ptflc}. The Mould-$R$ band luminosity remained mostly below our detection threshold up to 2 years before the outburst. After that, in what we call phase 1, it steadily increased in brightness by nearly two magnitudes between $-$2 and $-$1 years.  During phase 2, at the start of the summer 2014 ($-250$\,days) the progenitor was an additional magnitude brighter than in spring 2014, but faded by nearly a magnitude over the next three months. During phase 3, starting about 150\,days prior to the January 2015 outburst, the Mould-$R$ band flux reached its overall maximum of $\sim$20\,mag and then remained relatively steady, with some minor fluctuations.

CFHT data taken at about 3 months before eruption \citep{Dong15} agrees with iPTF observations (see insets in Figure \ref{fig:m31lc}). A single black body fit to the $g$ and $r$-bands SED shows that the temperature of the emission was 6300$^{+500}_{-500}$\,K with a radius of 55$^{+10}_{-10}$\,\Rsun, which indicates that the precursor emission became hotter than the progenitor and its photospheric radius increased. The lack of extensive simultaneous multi-band photometry during the whole precursor period makes it difficult to determine whether the optical luminosity is linked to the apparent expansion and cooling of the photosphere, similar to the behaviour displayed by other stellar mergers such as V1309\,Sco \citep{Pejcha2017ApJ} or M101-2015OT \citep{Blagorodnova17}, or to tidal heating \citep{MacLeod17}, where the atmosphere of the primary star is disturbed by the secondary. Like other stellar mergers, \objname also shows a smooth, monotonic brightening lasting about two years. After this, the light curve also shows a dip, followed by a short precursor, similar to the one observed for V838\,Mon \citep{Munari2002AA} or V1309\,Sco. This behaviour is likely too complex to be caused by tidal heating. A possible explanation is the loss of mass from the system through the the L2 point during the RLOF, as proposed by \citet{Pejcha2017ApJ}.

Mass loss progressively drains angular momentum from the binary, which causes the orbit to shrink. The outflowing gas can then form an expanding photosphere around the binary. \citet{Pejcha2017ApJ} showed that the slow rise of V1309\,Sco for thousands of orbits prior to merger could be explained by such dynamical mass loss. Assuming that the brightening in phase 1 of \objname{} is due to an expanding optically thick outflow, we would expect the luminosity to scale as $L \propto (\dot{P}/P)^2 \propto (\dot{M}/M)^2$, where $P$ is the period of the system.  As the mass loss increases, the outflow is expected to completely engulf the binary, thereby causing the luminosity to decrease (phase 2). The observed emission now comes from shock interactions created by the L2 outflow. Depending on the binary mass ratio, further mass loss from the system can either form an expanding photosphere, or be marginally bound to the binary and interact with it. Either of these scenarios leads to a gradual increase in the luminosity (phase 3), as the orbit shrinks further until the binary finally merges, leading to the outburst.

Though the general trend of the \objname light curve is similar to that of the V1309\,Sco, there are a few differences. The most prominent difference is that the pre-outburst brightening of V1309\,Sco smoothly transitioned into the outburst, with the photosphere expanding $\propto \dot{P}/P$, whereas for \objname the pre-outburst brightening reached a constant value before the main outburst. In this regard, \objname is more similar to V838\,Mon. Before its first eruption, V838\,Mon brightened by more than 3\,mags and plateaued for around 25 days \citep{Munari2002AA}. \citet{Tylenda05a} found that during this phase, the photospheric radius of V838\,Mon remained almost the same. This could occur if the mass ejected during the brightening is marginally bound, and falls back to interact with the binary \citep{Pejcha16b}. For outflows launched in synchronicity with the binary, a necessary, but not sufficient condition is the binary mass ratio $q$ to be $<$ 0.064 or $>$ 0.78.

\begin{figure}
\includegraphics[width=0.5\textwidth, angle=0]{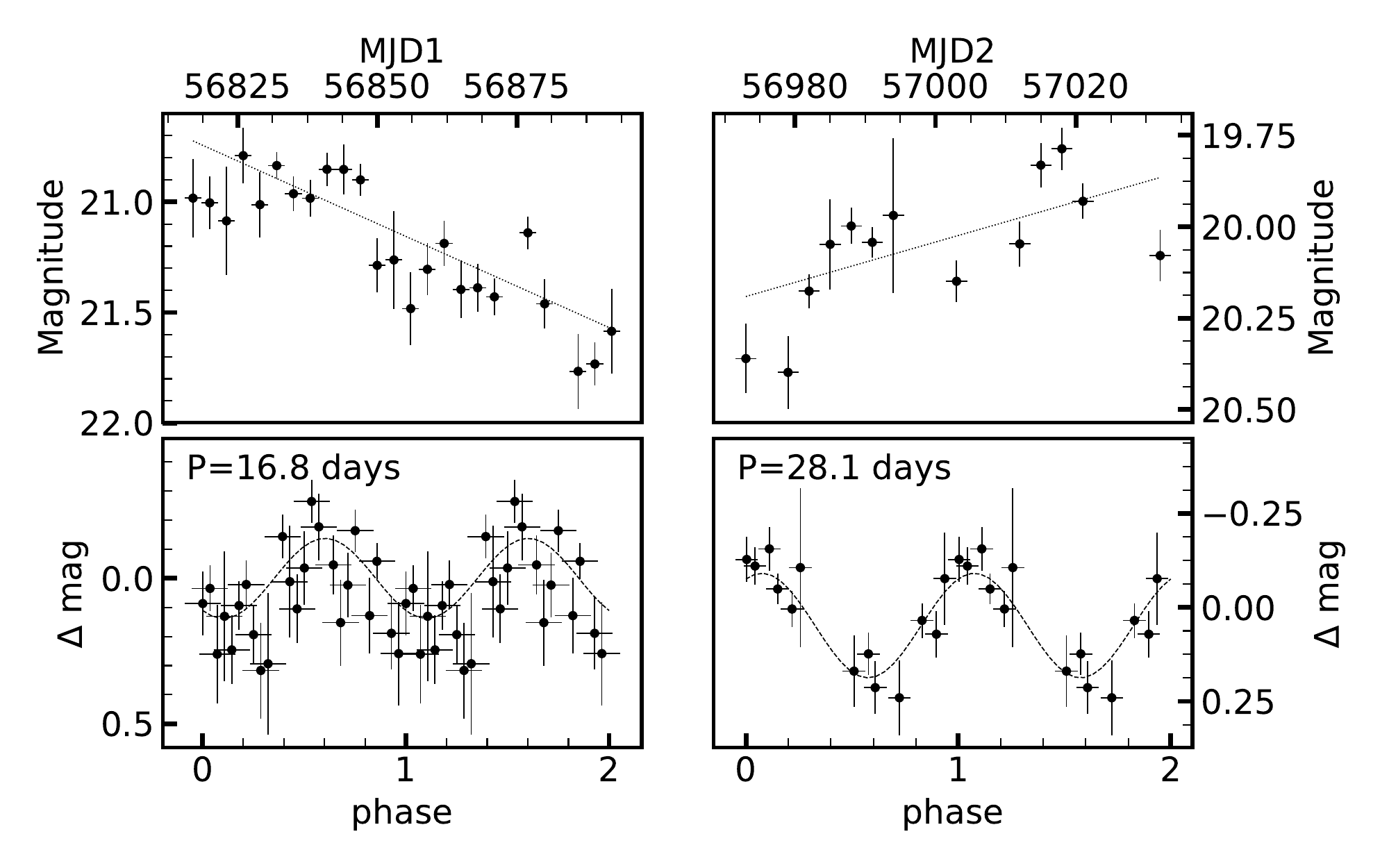}
\caption{Left: 3-day binned $R$-band light curve for phase 2 (top) and the folded light curve using the best average period of 16.8\,days (bottom). Right: 3-day binned light curve for phase 3  (top) and folded light curve with a period of 28.1\,days (bottom). The error bars in the X-axis represent the width of the bin. All panels show the best linear polynomial fit used to remove the continuum, as well as the model light curve for the corresponding period. \label{fig:lc_period}}
\end{figure}

Close inspection of phases 2 and 3 shows that the residuals seem to be correlated once the general trend is removed, as the scatter in our measurements is larger than the error bars (the uncertainties for phase 1 are too large for a reliable analysis). Similar characteristics were observed for V1309\,Sco \citep{Pejcha2017ApJ}, where the amplitude of the residuals oscillated by 0.3\,mag.  Deviations from the $(\dot{P}/P)^{2}$ rise were observed, but no period could be determined for the residuals. These residuals were interpreted as being due to the clumpiness and asymmetry of the mass lost from the binary. Motivated by the apparent correlation, we tested our light curve for periodicity in phase 2 (between MJD 56800$-$56895) and phase 3 (MJD 56960$-$57035). After removing a linear trend from our data, we used the Lomb-Scargle \citep{Lomb76,Scargle82} algorithm, to produce a periodogram using different binnings and offsets of the bins. For phase 2, none of our best fits exceeded a power of 0.5, indicating a low significance for any period. However, the results consistently show one best period around 17\,days (see Fig. \ref{fig:lc_period}).  Averaging all the periods for binnings from 1 to 4 days (weighted by their power) allows us to derive a possible period of $16.8 \pm  0.3$\,days. For phase 3, the power increases above 0.7. The average period for this phase is $28.1 \pm 1.4$\,days, nearly double than for phase 2.

If the period detected in phase 2 is real, its variability can be used to infer the characteristics of the binary system. \citet{MacLeod17} derived a mass of 3$-$5.5\,\Msun and a radius of 28$-$35\Rsun for the primary (depending on reddening). The period of a test mass at the surface of this binary is $13-7$ days depending on the mass and radius. However, if the primary is undergoing RLOF, the inferred radius corresponds to the Roche lobe of the primary. For our extinction value of $E(B-V)$=0.255, the primary has $M_1=5$\,\Msun and $R_1=$32\,\Rsun. The period for the binary for a mass ratio of $0.01 \leq q \leq 1$ ($q=M_2/M_1$) is in the range of 15$-$28\,days. 

Our first period of $\simeq 16.8$ days would be in agreement with RLOF scenario with $q \simeq 0.03$. However, an ellipsoidal variation would actually imply a period twice as large, which is outside the expected range we just derived. Our second period of $\simeq 28$\,days requires the binary to be of nearly equal mass. Ideally, earlier stages of the light curve would have provided an independent verification of the variability, but this is not possible due to the low SNR of the individual detections in our data. 

The detection of periodicity at later stages of the lightcurve is nevertheless surprising, as the binary is expected to be completely engulfed at this point. For example, an apparent increase in the period was observed in the early phases of the  V1309\,Sco light curve, when the phased light curve changed from a double hump profile to a single hump, effectively doubling the detected period from 0.7 to 1.4 days \citep{Tylenda11}. This doubling of the period was attributed to the asymmetric nature of the L2 mass loss, in that the ejected mass trailed around the binary and partly obscured the system. Provided the period for \objname has increased by nearly a factor of 2 from phase 2 to phase 3, it could possibly be a manifestation of the same phenomenon as in V1309\,Sco. It is also likely that the binary is in fact completely engulfed by the dust at this point, and the observed period is a result of interactions in the L2 mass outflow, as analyzed for the lightcurve of V1309\,Sco \citep{Pejcha14}.

\subsection{Post-Outburst Remnant Evolution} \label{sec:preoutburst}

The late time photometric evolution of \objname was monitored in the infrared by a combination of ground and space telescopes, as detailed in Section \ref{sec:photometry}. In order to analyze the SEDs of the transient, we fixed the times and values for the NIR observations, and linearly interpolated the \textit{Spitzer} and optical observations to those epochs. In total, we obtained the SED for seven post-peak epochs from at 43\,days up to $\sim$3.5\,years post peak.

Following the methods of \citet{Adams15} and \citet{Adams16,Adams17}, we modelled the SEDs with the dusty radiative transfer code \texttt{DUSTY} \citep{Ivezic97,Ivezic99,Elitzur01}. In these models, \texttt{DUSTY} is embedded in an Markov Chain Monte Carlo  (MCMC) wrapper to estimate the model parameters and their uncertainties. For the central source, we assume a black body radiation with a uniform prior on the temperature of 2500$\leq T_* \leq 20000$\,K. The dust is treated as a shell with density profile $\rho \propto r^{-2}$. We assume that the shell was launched on the date of discovery (2015 Jan 14th), when the object brightened significantly. The velocity of the shell, $v_{ej}$, is computed as the ratio between the outer radius and the time elapsed since its ejection. In our analysis, we fix the outer radius of the dust shell to be twice the inner dust radius, $R_{\rm{in}}$. The temperature of the dust at the inner radius is $T_*$ and at the outer radius is $T_{\rm{d}}$. We examine optical depth values in the visual band within the range $0.0 \leq \tau_V \leq 1000$.

For our analysis, we tried with both silicate and graphitic grains from \citet{Draine84}. For the dust grain sizes, we assume the standard MRN \citep{Mathis77} distribution function given by $n(a) \propto a^{-q}$, $a_{\mathrm{min}} < a < a_{\mathrm{max}}$ with q = 3.5, $a_{\mathrm{min}} = 0.005 \mu$m and $a_{\mathrm{max}} = 0.25 \mu$m. Finally, the reddening from the Milky Way and within the host galaxy was included as a fixed parameter with $E(B-V)$=0.255. 

One limitation of our analysis is the assumption of spherical symmetry. Outflows from stellar mergers are likely equatorial, due to previous episodes of mass loss in the binary orbital plane \citep{Metzger2017,MacLeodOstriker2018ApJ,Reichardt2019MNRAS}. Therefore, the estimated dust optical depth could be an over or underestimation depending on the viewing angle.

\begin{figure*}
\includegraphics[width=\textwidth]{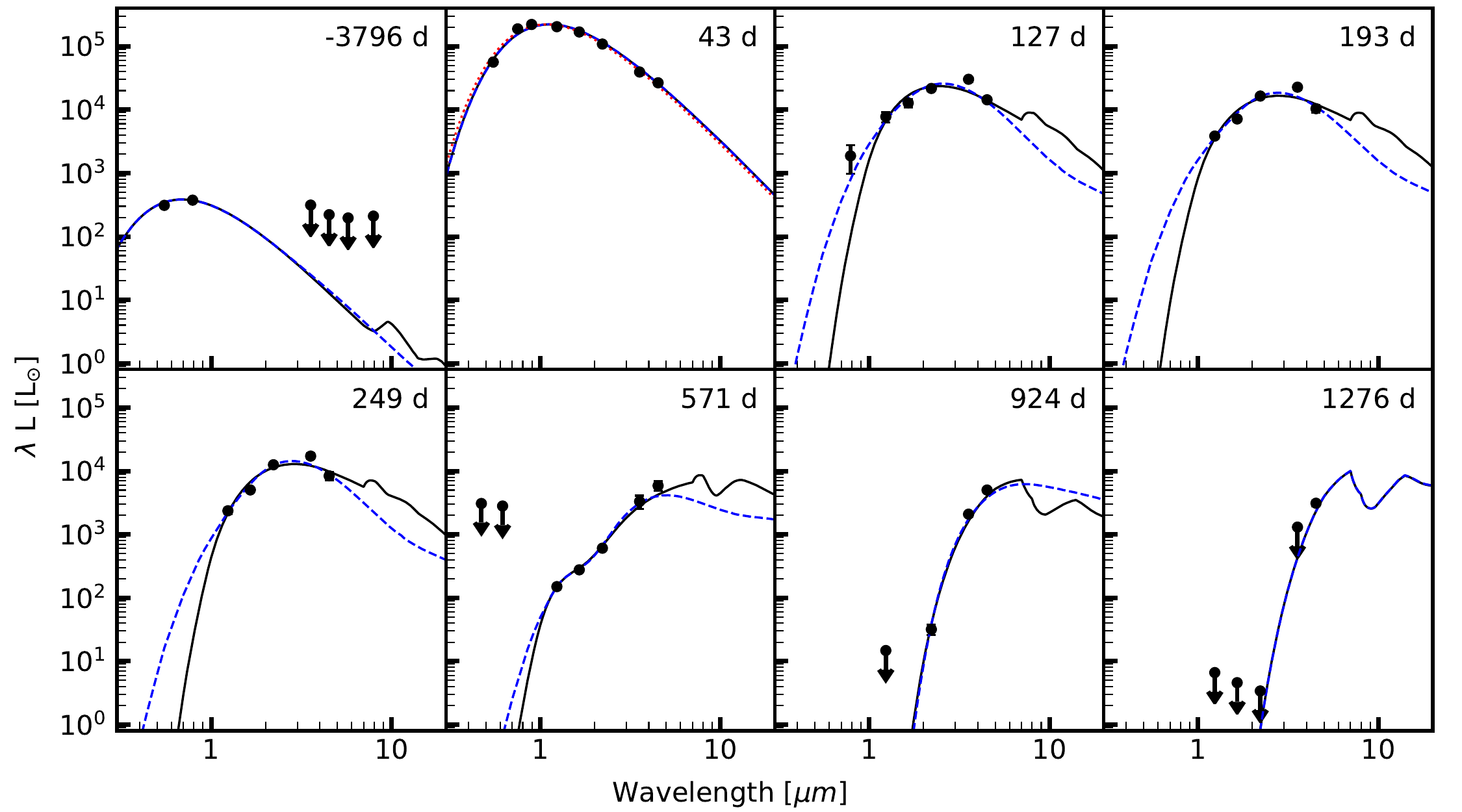}
\caption{Evolution of the SED of \objname with phases measured relative to MJD 57043.7. The SEDs in this Figure have been corrected for foreground extinction of $E(B-V)=0.255$. The black solid line shows the best fit silicate \textsc{dusty} model and the blue dashed line the best fit graphitic model. The red dotted line at 43\,days shows the contribution of a black body alone. \label{fig:seds}}
\end{figure*}

\begin{figure}
\includegraphics[width=0.5\textwidth]{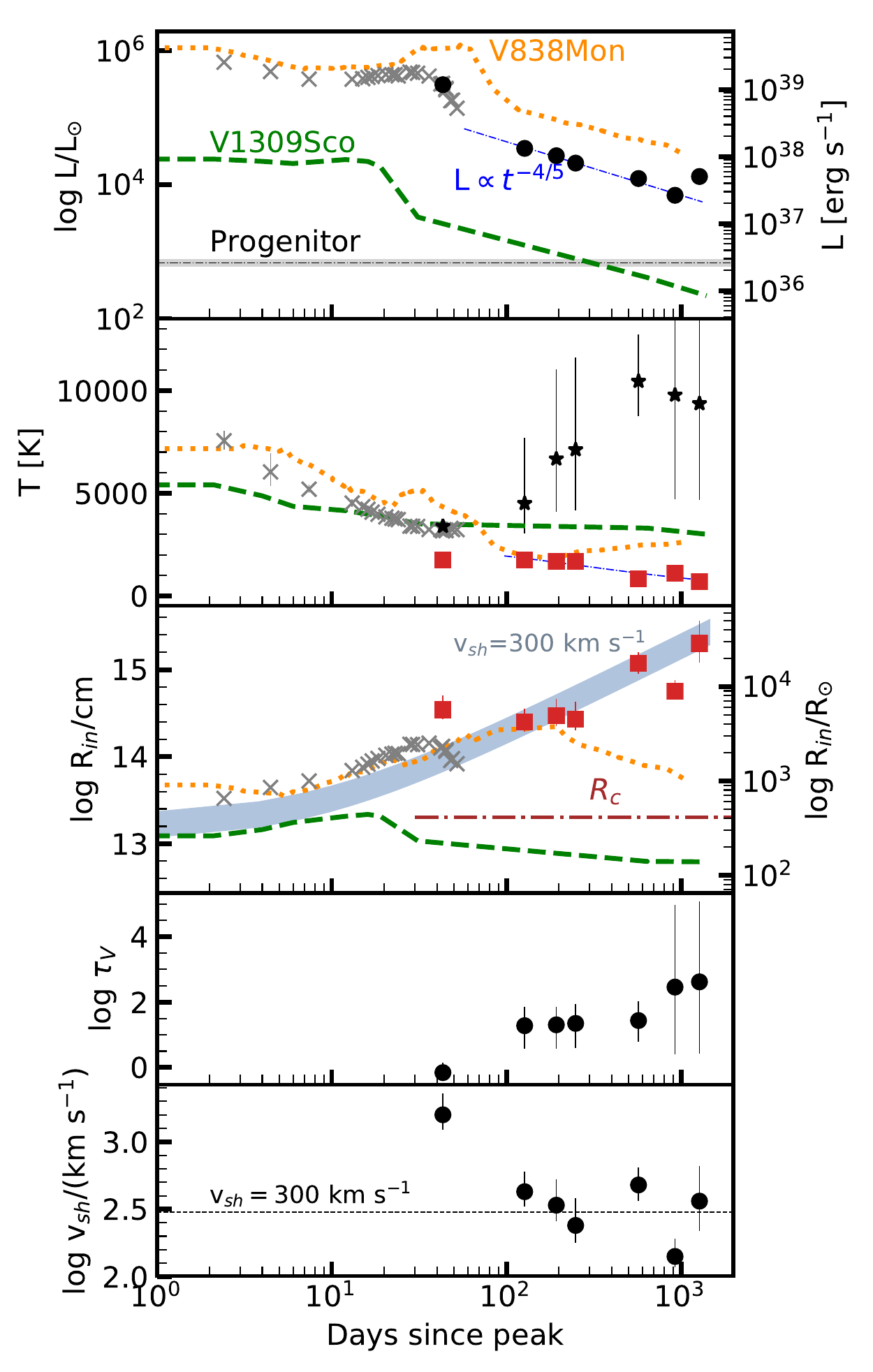}
\vspace{-0.5cm}
\caption{ The top panels show the bolometric luminosity of the system (black circles). The progenitor luminosity is marked with a dashed line. The blue line shows the expected gravitational contraction at fixed temperature (Eqn. \ref{eqn:lumdecay}). For context, we also include the evolution of the transient parameters derived from the optical and NIR SEDs by \citet{MacLeod17} (gray crosses). The coloured lines show the evolution for two Galactic stellar mergers \citep[V838\,Mon,V1309\,Sco;][]{Tylenda05a,Tylenda11,Tylenda2016}, where we  match the transients at the time of the first peak. The second panel shows the temperature for our late time analysis of the remnant star (black stars) and the dust shell (red squares), assuming silicate dust composition. The third panel shows the inner radius of the dust shell. The blue shaded area shows the inner and outer radius of a shell with uniform velocity that would have been ejected at discovery date. The brown line shows the dust condensation radius $R_c$ for the \objname progenitor. The bottom two panels show the optical depth and the shell expansion velocity.  \label{fig:dust_results}}
\end{figure}

The best fit models for both silicate and graphitic dust analysis are shown in Figure \ref{fig:seds}. The model parameters and their 1$\sigma$ uncertainties are provided in Table \ref{tab:sedmodels}, and depicted in Figure \ref{fig:dust_results}. Next we highlight the most important results of our analysis.

\begingroup
\renewcommand{\tabcolsep}{2pt}
\begin{table*}
\begin{center}
\begin{minipage}{16cm}
\caption{Posterior parameter from \textsc{dusty} MCMC models of the remnant SED.  The uncertainties give 1$\sigma$ levels.  $L_{*}$ is the bolometric luminosity of the source, $T_{*}$ is the intrinsic effective temperature of the input SED, $\tau_{V,\mathrm{tot}}$ is the optical depth of the shell in $V$-band, $R_{\mathrm{in}}$ is the inner radius of the dust shell, where the dust has temperature $T_{\mathrm{d}}$. $R_{\mathrm{out}}/R_{\mathrm{in}}$ is the thickness of the dust shell, $v_{\mathrm{ej}}$ is the velocity of the shell (assuming a constant expansion rate), $\chi^{2}$ is the fit of the model and $M{_\mathrm{ej}}$ the ejecta mass computed using Eqn. \ref{eq:ejmass}. For the analysis, we adopted the local $E(B-V)$=0.2 in M31 in addition to $E(B-V)=0.055$ for Galactic extinction.}
\label{tab:sedmodels}
\begin{tabular}{llrrrrrrrrrr}
\hline
\hline
{Dust} & {Date} & {Phase} & {log } & {$T_{*}$} & {$\tau_{V,\mathrm{tot}}$} &
{log } & 
{$T_{\mathrm{d}}$} &  {log } & {$\chi^{2}_{\mathrm{min}}$}  & $M{_\mathrm{ej}}$\\
 model & [UTC] & [d] & ($L_{*}/L_{\odot}$) & [K] &  & ($R_{\rm{in}}$/cm) & [K] & ($v_{\mathrm{ej}}$/km s$^{-1}$) & & [\Msun] \\ \hline
 silicate & 2004-08-16 & -3811  & $2.78^{+0.09}_{-0.04}$  & $5850^{+900}_{-660}$  & $0.8^{+0.8}_{-0.5}$  & $14.48^{+0.29}_{-0.29}$  & $440^{+220}_{-140}$  & $2.49^{+0.29}_{-0.29}$  & $1.8^{+2.4}_{-1.2}$ & 7.8$^{+33.5}_{-6.5} \times 10^{-6}$\\ 
 silicate & 2015-03-06 & 43  & $5.49^{+0.01}_{-0.01}$  & $3390^{+240}_{-130}$  & $0.7^{+0.8}_{-0.5}$  & $14.54^{+0.16}_{-0.11}$  & $1760^{+170}_{-250}$  & $3.2^{+0.16}_{-0.11}$  & $12.4^{+3.4}_{-2.3}$ & 1.0$^{+2.6}_{-0.8} \times 10^{-5}$\\ 
 silicate & 2015-05-29 & 127  & $4.54^{+0.03}_{-0.02}$  & $4500^{+3210}_{-1470}$  & $19.0^{+3.8}_{-5.1}$  & $14.4^{+0.15}_{-0.11}$  & $1740^{+160}_{-190}$  & $2.63^{+0.15}_{-0.11}$  & $26.1^{+3.5}_{-1.7}$ & 1.5$^{+0.8}_{-0.5} \times 10^{-4}$\\ 
 silicate & 2015-08-03 & 193  & $4.43^{+0.05}_{-0.03}$  & $6670^{+4360}_{-2560}$  & $20.2^{+3.6}_{-5.4}$  & $14.47^{+0.19}_{-0.12}$  & $1680^{+180}_{-180}$  & $2.53^{+0.19}_{-0.12}$  & $32.3^{+3.7}_{-1.9}$ & 2.2$^{+1.7}_{-0.8} \times 10^{-4}$\\ 
 silicate & 2015-09-28 & 249  & $4.32^{+0.05}_{-0.03}$  & $7120^{+4490}_{-2970}$  & $22.3^{+4.0}_{-5.6}$  & $14.43^{+0.2}_{-0.13}$  & $1680^{+180}_{-200}$  & $2.38^{+0.2}_{-0.13}$  & $28.4^{+3.7}_{-2.0}$ & 2.0$^{+1.9}_{-0.7} \times 10^{-4}$\\ 
 silicate & 2016-08-15 & 571  & $4.09^{+0.11}_{-0.11}$  & $10450^{+2280}_{-1700}$  & $27.3^{+3.9}_{-4.4}$  & $15.07^{+0.13}_{-0.12}$  & $840^{+50}_{-50}$  & $2.68^{+0.13}_{-0.12}$  & $3.7^{+3.4}_{-2.0}$ & 4.7$^{+3.7}_{-2.0} \times 10^{-3}$\\ 
 silicate & 2017-08-03 & 924  & $3.84^{+0.08}_{-0.06}$  & $9780^{+6200}_{-5090}$  & $287.2^{+324.7}_{-113.2}$  & $14.75^{+0.13}_{-0.08}$  & $1110^{+210}_{-190}$  & $2.15^{+0.13}_{-0.08}$  & $3.1^{+2.1}_{-0.7}$ &  1.3$^{+0.9}_{-0.5} \times 10^{-2}$ \\ 
 silicate & 2018-07-21 & 1276  & $4.12^{+0.26}_{-0.19}$  & $9360^{+6340}_{-4690}$  & $415.2^{+280.7}_{-157.8}$  & $15.3^{+0.26}_{-0.22}$  & $700^{+150}_{-120}$  & $2.56^{+0.26}_{-0.22}$  & $0.7^{+1.4}_{-0.4}$ & 2.1$^{+5.3}_{-1.4} \times 10^{-1}$ \\ 
 \hline
 graphite & 2004-08-16 & -3811  & $2.79^{+0.08}_{-0.05}$  & $5840^{+780}_{-600}$  & $0.3^{+0.3}_{-0.2}$  & $14.44^{+0.36}_{-0.34}$  & $760^{+270}_{-210}$  & $2.44^{+0.36}_{-0.34}$  & $2.5^{+3.1}_{-1.6}$  & 2.7$^{+12.9}_{-2.2} \times 10^{-6}$ \\ 
 graphite & 2015-03-06 & 43  & $5.48^{+0.01}_{-0.01}$  & $3410^{+190}_{-130}$  & $0.2^{+0.3}_{-0.2}$  & $14.53^{+0.13}_{-0.08}$  & $1820^{+120}_{-190}$  & $3.19^{+0.13}_{-0.08}$  & $13.1^{+3.5}_{-2.3}$ & 3.9$^{+5.8}_{-2.8} \times 10^{-6}$ \\ 
 graphite & 2015-05-29 & 127  & $4.5^{+0.02}_{-0.01}$  & $5600^{+3810}_{-2110}$  & $5.6^{+1.7}_{-2.4}$  & $14.69^{+0.16}_{-0.13}$  & $1300^{+140}_{-110}$  & $2.92^{+0.16}_{-0.13}$  & $20.9^{+3.2}_{-2.0}$ &1.6$^{+0.5}_{-0.4} \times 10^{-4}$\\ 
 graphite & 2015-08-03 & 193  & $4.36^{+0.03}_{-0.02}$  & $6970^{+3930}_{-2520}$  & $5.5^{+2.1}_{-2.8}$  & $14.75^{+0.18}_{-0.14}$  & $1220^{+100}_{-90}$  & $2.81^{+0.18}_{-0.14}$  & $21.8^{+3.1}_{-1.9}$ & 2.0$^{+0.7}_{-0.5} \times 10^{-4}$ \\ 
 graphite & 2015-09-28 & 249  & $4.25^{+0.04}_{-0.02}$  & $7900^{+5070}_{-3320}$  & $5.7^{+2.8}_{-3.6}$  & $14.72^{+0.22}_{-0.16}$  & $1200^{+110}_{-100}$  & $2.68^{+0.22}_{-0.16}$  & $17.4^{+2.8}_{-1.7}$ & 1.8$^{+0.6}_{-0.5} \times 10^{-4}$ \\ 
 graphite & 2016-08-15 & 571  & $3.83^{+0.09}_{-0.09}$  & $7320^{+1070}_{-910}$  & $11.3^{+1.5}_{-1.5}$  & $15.24^{+0.12}_{-0.12}$  & $670^{+40}_{-40}$  & $2.84^{+0.12}_{-0.12}$  & $5.2^{+3.4}_{-2.0}$ & 4.2$^{+3.5}_{-1.9} \times 10^{-3}$ \\ 
 graphite & 2017-08-03 & 924  & $3.97^{+0.15}_{-0.16}$  & $9580^{+5420}_{-4490}$  & $68.2^{+32.8}_{-8.8}$  & $15.0^{+0.35}_{-0.44}$  & $890^{+350}_{-170}$  & $2.4^{+0.35}_{-0.44}$  & $2.4^{+3.1}_{-1.7}$ & 9.5$^{+30.6}_{-7.0} \times 10^{-3}$\\ 
 graphite & 2018-07-21 & 1276  & $4.1^{+0.26}_{-0.2}$  & $6980^{+6900}_{-3100}$  & $441.1^{+289.3}_{-184.6}$  & $15.28^{+0.26}_{-0.22}$  & $680^{+130}_{-110}$  & $2.54^{+0.26}_{-0.22}$  & $0.8^{+1.5}_{-0.5}$ & 2.0$^{+5.3}_{-1.3} \times 10^{-1}$\\ 
 \hline
\end{tabular}
\end{minipage}
\end{center}
\end{table*}
\endgroup

The \texttt{DUSTY} analysis of the progenitor SED obtained at $-$3795\,days (10.4\,years prior to the outburst) favours optically thin dust. This is consistent with our analysis in Section \ref{sec:preoutburst}, where for a dust temperature of $T_d \sim 400$\,K the total dust mass in the system was below 10$^{-7}$\,\Msun. 

Our earliest full optical to MIR SED is located towards the end of the $r$-band plateau at 43\,days. At this stage, the transient is well-fit by a single component $\simeq$3400\,K blackbody and a radius of 1700\,\Rsun, which is consistent with the size of the ejecta if assuming that it expanded at a uniform velocity of 300\,\kms. Albeit the evidence of dust at this phase is small (see the 43\,days panel in Figure \ref{fig:seds}), we discuss the possible implications. If dust is present, it is optically thin and located far from the stellar photosphere.  Our spectrum taken at +47.4\,days already shows signs for \io{TiO}{} absorption, in agreement with a cold stellar temperature. The shell has a dust temperature of $\simeq$1750\,K, which would allow the dust condensation process to take place. If ejected around the discovery date, the large extent of this shell implies a much faster velocity than the velocity of the photosphere, with 1580$^{+700}_{-360}$\,\kms. The most likely scenario is that this shell was already formed during the years or months preceding the eruption. For example, if the gas travelled from the system at the escape velocity of $\sim$250\,\kms \citep{MacLeod17}, a thin shell could have been ejected $\sim$160\,days before the outburst peak, which coincides with the luminosity increase in the transient's light curve (phase 3).

By 127\,days, the emission in the optical has almost faded below detection, as the dust shell has become optically thick ($\tau_{V} \simeq 19$) and has started to reprocess most of the radiation from the obscured remnant. The temperature of the star appears slightly hotter than during the plateau, with $T_* \simeq4500$\,K. The radius of this shell is $\sim$4000\,\Rsun. Within the errors, this is consistent with the estimates for the previous phase when the shell had to be optically thin. The estimated temperature of this shell is also similar, $T_d \sim 1700$\,K, which is close to the threshold for the condensation of O-rich minerals \citep{Lodders2003ApJ}.

A couple of months later, at 193\,days and 249\,days, we observe a continuous decrease in the bolometric luminosity of the source. Although the temperature of the star increases to $T_* \sim 7000$\,K, the temperature and the inner radius of the dust shell remains relatively constant. Over the first year of observations, the optical depth of the shell also remains stable.

A major change in the system is observed at 571\,days ($\sim$1.5\,years post-outburst). At this time, the SED has clearly developed two distinct emission components. While the temperature of the central star has risen even further to $T_*\sim 10^4$\,K, the temperature of the dust has cooled to $\sim$800\,K. The shell also appears four times larger than before with $R_{\rm{in}}\sim 1.6\times 10^4$\,\Rsun. 

After this phase, our last two epochs taken at 2.5 and 3.5\,years show that the star possibly maintains a temperature of $\sim10^4$\,K. The lack of optical data makes this temperature mostly dependent on our prior. The dust temperature at 2.5 years appears slightly warmer than before, $\sim$1100\,K and its internal radius shrinks to $R_{\rm{in}}\sim 8\times 10^3$\,\Rsun. However, by 3.5\,years it cools again to 700\,K and expands to its largest value of $R_{\rm{in}}\sim 3\times 10^4$\,\Rsun. The optical depth for these last two epochs is $\tau_{V} > 200$. Although our analysis for this phase is mostly based on non-detections, we suggest a possible final increase in the total luminosity of the source.

To summarize our analysis, we see that dust formation occurs in three main episodes. First, there is weak evidence for an initial ejection of gas at $\sim$160\,days before the outburst, forming an optically thin shell of warm dust present during the post peak plateau.

The second episode takes place during the first 1.5\,years after the outburst. The shell cools slowly, forming a moderate amount of dust ($\tau_V \simeq 20$). Its inner radius is mostly consistent with the location of matter ejected at $v_{\rm{sh}}\sim 300$\,\kms near the discovery date (see Figure \ref{fig:dust_results}). 

The final episode includes epochs later than 2.5\,years. They are characterized by large optical depths, with values $\tau_V >$200. In the first epoch, there is an apparent shrinkage of the inner radius. This measurement is not in agreement with the expanding shell trend, which may be a limitation of our model due to the limited data on the source's SED. However, if the radius contraction is real, we attribute it to an increase in the dust opacity, perhaps due to the formation of smaller dust grains deeper in the ejecta (see \citet{Iaconi2020arXiv}) or to an increase in the dust temperature due to shocks. This argument is supported by our last epoch at 3.5\,years: once the dust has cooled again, the radius seems to return to the trend of an expanding shell. Progressive cooling of the shell allows further dust condensation, increasing its mass up to an order of magnitude (see Table \ref{tab:sedmodels}).

\subsection{Estimate of the ejecta mass}

For a shell with total ejecta mass $M_{ej}$ and radius $R$, the optical depth in the visual band is defined as

\begin{equation}
   \tau_V = \frac{\kappa_V M_{\mathrm{ej}}}{4 \pi R^2} \approx  3.27\times10^6 \left(\frac{\kappa_V}{100\rm{cm}^2/{g}} \right) 
   \left(\frac{M_{\mathrm{ej}}}{M_{\odot}}\right)
   \left(\frac{1000R_{\odot}}{R} \right)^2
\end{equation}
Provided our models fit for $\tau_V$ and the shell inner radius, we can estimate the total ejected mass as
\begin{equation}
\label{eq:ejmass}
\frac{M_{\mathrm{ej}}}{M_{\odot}} \approx 3.06 \times 10^{-7}   \left(\frac{100 \rm{cm}^2/\rm{g}}{\kappa_V}\right)\left(\frac{R}{1000 R_{\odot}}\right)^2 \tau_V
\end{equation}
Given that $\kappa_V \sim 50-100$\,cm$^2$g$^{-1}$ for a typical dust-to-gas ratio \citep{Ossenkopf1994}, we derive the mass of the shell at different epochs (see Table \ref{tab:sedmodels}). For our last epoch SED and adopting $\kappa_V$=50\,cm$^2$/g we derive $M_{\rm{ej}}=$0.21$^{0.53}_{-0.13}$\,\Msun, which is consistent with the estimates by \citet{Williams2015} and \citet{MacLeod17}. Nevertheless, this value is likely only a lower limit, as simulations have shown that dust formation continues in the inner regions of the ejecta many years after the outburst has faded \citep{Iaconi2020arXiv}.

This estimate brings some tension with the models of \cite{Metzger2017}. In their shock powered scenario, the same luminosity would require a substantially lower ejecta mass than the 0.3\,\Msun derived for a recombination scenario \citep{MacLeod17}. For shell velocities in the range of $100-300$\,\kms the ejecta mass would drop by a factor of $\sim (2 \rm{Ryd}/(m_p v_{\rm{sh}}^2)) \sim 3-30$, requiring at most 0.1\,\Msun. However, assumptions about the gas opacity can introduce large systematic errors in the mass estimates and remove the discrepancy.

\section{Discussion} \label{sec:discussion}

\subsection{The origin of the precursor emission} \label{sec:precursor}

\begin{figure}
\includegraphics[width=0.5\textwidth]{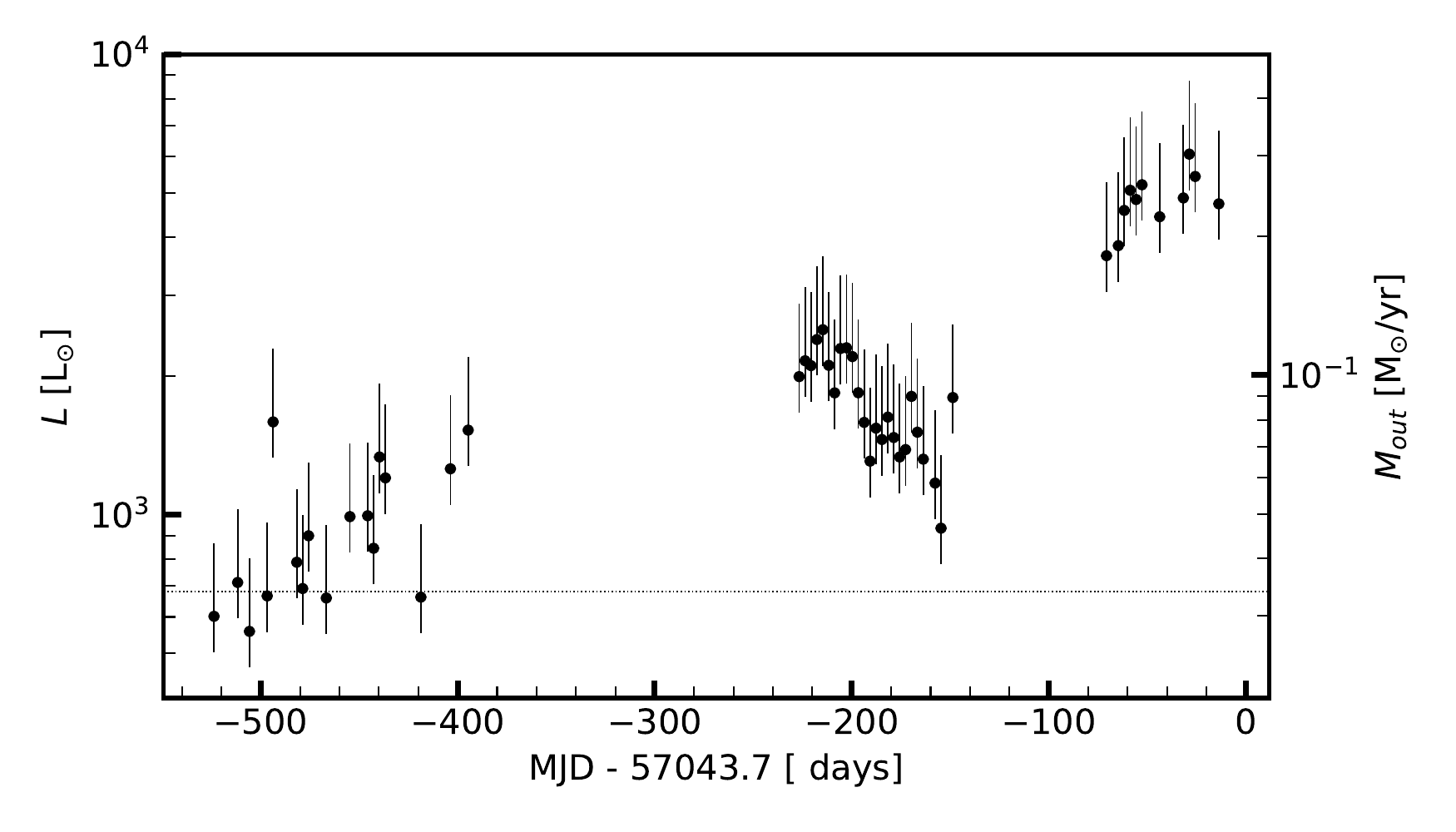}
\caption{Bolometric luminosity of the precursor emission and mass loss rate corresponding to an isothermal outflow. The line shows the luminosity of the progenitor.}
\label{fig:precursor_mdot}
\end{figure}

Our observations show an increase in the luminosity prior to the primary transient. One possible origin for the brightening is the dissipation of energy by shocks within the spiral stream that forms around the binary once the gas has left the L2 point. In order to infer the mass loss required to power this emission, we first reconstruct the total luminosity of the system (see Figure \ref{fig:precursor_mdot}). We assume a black body emission at the estimated temperature of the progenitor star ($T_{\rm{out}}=4300$\,K). Scaling the synthetic photometry of the black body to the observed flux in $r$-band allows us to estimate the total bolometric luminosity of the system. We vary $T_{\rm{out}}$ between 3500 and 8000\,K, to illustrate the effect of the temperature uncertainty.

Following \citet{Pejcha16b}, the luminosity of an optically thin outflow powered by the dissipation of energy from shocks within the stream is

\begin{equation}
    L_{\rm{thin}} \sim \frac{1}{2} \dot{M} \Delta v^2
\end{equation}
where we adopt $\Delta v/v_{\rm{esc}}=0.5/\eta$ \,with $\eta=8$. When matched to the observed luminosity, the resulting mass loss rate is shown in the left axis in Figure \ref{fig:precursor_mdot}. The integrated mass loss within this period is $M_{\rm{out}}=1.4^{+0.6}_{-0.2} \times 10^{-1}$\,\Msun.

For an optically thick outflow, not all the energy is radiated, as part goes into the adiabatic expansion of the gas. Numerical results presented in \citet{Pejcha16b} showed that the maximum asymptotic luminosity of the outflow has a weak dependence on the vertical optical depth of the flow, $\tau_{z,c}$ with

\begin{equation}
    L_{\rm{thick,max}} \sim L_{\rm{thin}} \,\rm{max} (1, \tau_{z,c})^{-0.2}.
\end{equation}
Hence, if the temperature of the outflow is maintained constant, we should interpret our mass loss estimates as lower limits. However, a lower value could still be compatible with the observed luminosity. For example, an increase in the temperature at which matter is ejected from L2 can drive the spiral stream to becomes wider, causing the collisions to occur closer to the binary and increasing the radiative efficieny of the outflow \citep{Pejcha16,Pejcha2017ApJ}. Alternatively, if the outflow is bound to the system, it will form a circumbinary torus \citep{MacLeodOstrikerStone2018ApJ}, which can dissipate energy more efficiently than the thin disk approach.

\subsection{Late time evolution of the remnant} \label{sec:latetime}

The luminosity evolution of \objname at late times (after the termination of the plateau) can be modelled as gravitational contraction of an inflated envelope surrounding the remnant. Similar explanations have been given for the post-outburst luminosity declines observed in V4332\,Sgr and V838\,Mon. \citet{Tylenda05b} models the photometric evolution assuming gravitational contraction of an envelope with binding energy
\begin{equation}
E_{\mathrm{g}}=-\frac{GM_{*}M_{\mathrm{e}}}{(R_{*}R)^{1/2}}
\end{equation}
where $M_{*}$ is the mass of the remnant, $M_{\mathrm{e}}$ is the mass of the envelope,  $R_{*}$ is the inner radius of the envelope and $R$ is the outer radius of the envelope.  If the luminosity is primarily from gravitational contraction, then
\begin{equation}
L=-\frac{dE_{\mathrm{g}}}{dt}=\frac{GM_{*}M_{\mathrm{e}}}{2R_{*}^{1/2}}\frac{1}{R^{3/2}}\frac{dR}{dt}
\end{equation}
and since $L=4\pi R^{2}\sigma T^{4}_{\mathrm{eff}}$,
\begin{equation}
\label{eq:dr}
\frac{dR}{R^{7/2}}=-\frac{8\pi R_{*}^{1/2}\sigma T^{4}_{\mathrm{eff}}}{GM_{*}M_{\mathrm{e}}}dt
\end{equation}
For sake of simplicity, we will assume constant $T_{\mathrm{eff}}$ even though our SED analysis shows a possible increase in the temperature. In these phases our estimate of $T_*$ is strongly influenced by our priors and we consider our approximation acceptable for illustrative purposes. With $T_{\mathrm{eff}}$ fixed, Eqn \ref{eq:dr} can be integrated to

\begin{equation}
\label{eqn:lumdecay}
\frac{1}{R^{5/2}}=\frac{20\pi R_{*}^{1/2}\sigma T^{4}_{\mathrm{eff}}}{GM_{*}M_{\mathrm{e}}}t+\frac{1}{R_{0}^{5/2}}
\end{equation}
where $R_{0}$ is the envelope radius at the start of the contraction at ($t=0$).  
In this simplified model, the late-time luminosity evolution of a gravitationally settling envelope should be $L\propto t^{-4/5}$, which is plotted in the upper panel of Figure \ref{fig:dust_results}. This trend, similar to the one for V838\,Mon and V1309\,Sco, is observed until 2.5\,years after peak. After that, the object seems to increase in luminosity possibly due to shocks in the ejecta.

\subsection{The effect of dust driven winds} \label{sec:winds}

Dust-driven winds are one of the main mechanisms in removing the extended envelopes in asymptotic giant-branch (AGB) stars. The low-temperature and high density environments generated by stellar pulsations are suitable sites for dust growth. Because of the coupling between dust and gas, the radiation pressure from the star on dust grains provides an efficient way to accelerate the outermost layers to super-sonic velocities. 

In the case of stellar mergers or successful CE remnants, the envelope has an expanding unbound component, but there is also a significant amount of gas which is still gravitationally bound to the system \citep{Pejcha16b,MacLeodOstrikerStone2018ApJ}. If this gas is located beyond the dust condensation radius $R_C$, the dust grains will be able to form and grow. The radiation pressure on the bound gas will eventually accelerate the grains to super-sonic speeds and allow gas to escape from the system.

Following \citet{Glanz18}, we define the dust condensation radius using the properties of the primary star before the start of the CE
\begin{equation}
    R_C = 0.5^{2/5} \left( \frac{T_C}{T_*} \right)^{-2} R_*.
\end{equation}
Assuming the progenitor parameters ($T_*$, $R_*$) derived in Section \ref{sec:progenitor_dust}, a condensation temperature of $T_C=1500$\,K will be reached at $R_C \simeq 290$\,\Rsun. During the outburst, this radius is well within the photosphere of the nova, which is located at $\sim 10^4$\,\Rsun, as shown in Figure \ref{fig:dust_results}. 

Although a detailed analysis of the dust-driven outflows for \objname{} is outside of the scope of this work (see \citet{Iaconi2020arXiv} for an analysis of dust formation in CE ejecta), we would like to highlight that this process may provide the long-sought channel for unbinding large part of the gas from binary systems after the spiral-in and common-envelope ejection had taken place. 

\section{Summary and conclusions} \label{sec:conclusions}

In this work we have made an extended observational study of the stellar merger \objname{}, from 5\,years prior the peak, to 5\,years after the peak. The main conclusions of our study are:

\begin{itemize}
    \item The progenitor of \objname did not have a detectable dust emission component, which sets an upper limit on the dust masses between $10^{-6.4}$ and $10^{-9.7}$\,\Msun for optically thin dust with temperatures of 250\,K and 1500\,K respectively.
    
    \item The system began to brighten 2\,years before peak. At about $\sim-$200\,days, the light curve dimmed by 1\,mag, but the brightening resumed shortly afterwards. This complex luminosity evolution is best explained by the RLOF of the primary and mass loss from L2 point. Assuming the observed emission is mainly powered by shocks within the self-intersecting stream, the total mass loss during the precursor phase is $\geq 0.14$\,\Msun.
    
    \item The light curve of the precursor may show a period of 16.8$\pm$0.3\,days in the dimming phase and 28.1$\pm$1.4\,days  in the brightening phase. Although the significance of these period detections is low, we tentatively attribute this periodicity to the characteristics of the outflow from the binary, analogous to the trend observed in V1309\,Sco. 
    
    \item The spectroscopic evolution of the merger initially shows a 5000\,K continuum with a superposed forest of narrow absorption lines from low ionization elements. After peak, the temperature of the continuum quickly drops, and strong molecular absorption features of \ion{TiO}{} and \ion{VO}{} appear at +47\,days. By +140\,days, the object has no detectable emission blueward of 8000\,\AA\, and its spectrum resembles a Mira-like M9e type star.
    
    \item Before peak, the spectrum shows an \halpha emission component centred at the rest-frame velocity. The line temporarily disappears during the plateau phase, and then reappears with a blueshift of $\sim$300\,\kms. The redshifted component is likely obscured by dust formed in the eruption.
    
    \item During the post-peak light curve plateau, the SED suggests the existence of an optically thin dust shell located outside the expanding photosphere of the remnant star. For gas at the system's escape velocity, the shell would have been ejected about half a year before the nova event, coinciding with the last brightening episode in the precursor emission.
    
    \item After the end of the plateau, the luminosity decreases at a rate consistent with a gravitationally settling giant star. While the merger remnant may become hotter with time, the dust shell cools down from $\sim$1700\,K to $\sim700$\,K.
    
    \item At late times, the ejecta forms an optically thick ($\tau_V \sim 400$) shell that totally obscures the stellar remnant in optical and NIR wavelengths. The shell has an estimated mass of $M_{\rm{ej}}\sim 0.2$\,\Msun , in agreement with previous works. Its inner radius is located beyond the dust condensation radius, allowing radiation pressure to further accelerate this dusty matter.
\end{itemize}

The data and analysis presented in this work shows the importance of late time follow-up for a stellar merger in the infrared, revealing the fast dust formation mechanisms, which quickly obscure the nova remnant. Due to its large optical depth, the late-time merger remnant is undetectable in optical and near-infrared wavelenghts, although still visible in the mid-infrared.
Some of the SPRITEs (eSPecially Red Intermediate-luminosity Transient Events) reported by the SPIRITS survey \citep{Kasliwal2017ApJ,Jencson2019ApJ} show similar characteristics to \objname{} at late times, and may also be stellar mergers. The progressive evolution of the object's emission into longer wavelengths makes it an exciting target for the James Webb Space Telescope (JWST).

\section*{Acknowledgements}
\begin{small}
N. B. would like to thank O. Pejcha, M. MacLeod, B. Metzer, O. De Marco and N. Soker for useful discussions, A. Kurtenkov and S. C. Williams for making available the spectra of \objname, A. Pastorello for data on AT2017fjs, and T. Kaminski for the spectrum of V1309Sco. To M. Fraser, D. Perley, and R. M. Wagner for observations and data reduction, and to T. Szalai for the dust emission models. This work is part of the research programme VENI, with project number 016.192.277, which is (partly) financed by the Netherlands Organisation for Scientific Research (NWO). SK acknowledges the financial support of the Polish National Science Center (NCN) through the OPUS grant 2018/31/B/ST9/00334. P.E.N. acknowledges support from the DOE under grant DE-AC02-05CH11231, Analytical Modeling for Extreme-Scale Computing Environments. CSK is supported by NSF grants AST-1908570 and AST-1814440. This research benefited from interactions with Natasha Ivanova, Brian Metzger and Lars Bildsten that were funded by the Gordon and Betty Moore Foundation through Grant GBMF5076.
The Intermediate Palomar Transient Factory project is a scientific collaboration among the California Institute of Technology, Los Alamos National Laboratory, the University of Wisconsin, Milwaukee, the Oskar Klein Center, the Weizmann Institute of Science, the TANGO Program of the University System of Taiwan, and the Kavli Institute for the Physics and Mathematics of the Universe. The WHT spectrum was taken under program (2014B/P29). This research used resources of the National Energy Research Scientific Computing Center, a DOE Office of Science User Facility supported by the Office of Science of the U.S. Department of Energy under Contract No. DE-AC02-05CH11231.

Some of the data presented herein were obtained at the W.M. Keck Observatory, which is operated as a scientific partnership among the California Institute of Technology, the University of California and the National Aeronautics and Space Administration. The Observatory was made possible by the generous financial support of the W.M. Keck Foundation.
The authors wish to recognize and acknowledge the very significant cultural role and reverence that the summit of Maunakea has always had within the indigenous Hawaiian community.  We are most fortunate to have the opportunity to conduct observations from this mountain. 
This work is based in part on observations made with the Large Binocular Telescope. The LBT is an international collaboration among institutions in the United States, Italy and Germany. The LBT Corporation partners are: The University of Arizona on behalf of the Arizona university system; Istituto Nazionale di Astrofisica, Italy; LBT Beteiligungsge- sellschaft, Germany, representing the Max Planck Society, the Astrophysical Institute Potsdam, and Heidelberg University; The Ohio State University; The Research Corporation, on behalf of The University of Notre Dame, University of Minnesota and University of Virginia.
Part of this research was carried out at the Jet Propulsion Laboratory, California Institute of Technology, under a contract with the National Aeronautics and Space Administration.
We acknowledge Telescope Access Program (TAP) funded by the NAOC, CAS, and the Special Fund for Astronomy from the Ministry of Finance.
This work is based in part on observations made with the Spitzer Space Telescope, which is operated by the Jet Propulsion Laboratory, California Institute of Technology under a contract with NASA.
This research made use of Astropy,\footnote{http://www.astropy.org} a community-developed core Python package for Astronomy \citep{astropy:2013, astropy:2018}.
\end{small}

\textbf{Data availability: }
The photometric data and analysis results underlying this article are available in the article and in its online supplementary material.
The spectroscopic data underlying this article are available in the online repository \texttt{WiseRep} \citep{YaronGal-Yam2012} at \url{https://wiserep.weizmann.ac.il}, and can be accessed with the id ``iPTF15t''.

\bibliography{myreferences}

\begin{thebibliography}{}
\makeatletter
\relax
\def\mn@urlcharsother{\let\do\@makeother \do\$\do\&\do\#\do\^\do\_\do\%\do\~}
\def\mn@doi{\begingroup\mn@urlcharsother \@ifnextchar [ {\mn@doi@}
  {\mn@doi@[]}}
\def\mn@doi@[#1]#2{\def\@tempa{#1}\ifx\@tempa\@empty \href
  {http://dx.doi.org/#2} {doi:#2}\else \href {http://dx.doi.org/#2} {#1}\fi
  \endgroup}
\def\mn@eprint#1#2{\mn@eprint@#1:#2::\@nil}
\def\mn@eprint@arXiv#1{\href {http://arxiv.org/abs/#1} {{\tt arXiv:#1}}}
\def\mn@eprint@dblp#1{\href {http://dblp.uni-trier.de/rec/bibtex/#1.xml}
  {dblp:#1}}
\def\mn@eprint@#1:#2:#3:#4\@nil{\def\@tempa {#1}\def\@tempb {#2}\def\@tempc
  {#3}\ifx \@tempc \@empty \let \@tempc \@tempb \let \@tempb \@tempa \fi \ifx
  \@tempb \@empty \def\@tempb {arXiv}\fi \@ifundefined
  {mn@eprint@\@tempb}{\@tempb:\@tempc}{\expandafter \expandafter \csname
  mn@eprint@\@tempb\endcsname \expandafter{\@tempc}}}

\bibitem[\protect\citeauthoryear{{Adams} \& {Kochanek}}{{Adams} \&
  {Kochanek}}{2015}]{Adams15}
{Adams} S.~M.,  {Kochanek} C.~S.,  2015, \mn@doi [\mnras]
  {10.1093/mnras/stv1409}, \href
  {http://adsabs.harvard.edu/abs/2015MNRAS.452.2195A} {452, 2195}

\bibitem[\protect\citeauthoryear{{Adams}, {Kochanek}, {Dong}  \&
  {Wagner}}{{Adams} et~al.}{2015a}]{ATel7468}
{Adams} S.,  {Kochanek} C.~S.,  {Dong} S.,   {Wagner} R.~M.,  2015a, The
  Astronomer's Telegram, \href
  {http://adsabs.harvard.edu/abs/2015ATel.7468....1A} {7468}

\bibitem[\protect\citeauthoryear{{Adams}, {Kochanek}, {Dong}  \&
  {Wagner}}{{Adams} et~al.}{2015b}]{ATel7485}
{Adams} S.,  {Kochanek} C.~S.,  {Dong} S.,   {Wagner} R.~M.,  2015b, The
  Astronomer's Telegram, \href
  {https://ui.adsabs.harvard.edu/abs/2015ATel.7485....1A} {7485, 1}

\bibitem[\protect\citeauthoryear{{Adams}, {Kochanek}, {Prieto}, {Dai},
  {Shappee}  \& {Stanek}}{{Adams} et~al.}{2016}]{Adams16}
{Adams} S.~M.,  {Kochanek} C.~S.,  {Prieto} J.~L.,  {Dai} X.,  {Shappee} B.~J.,
    {Stanek} K.~Z.,  2016, \mn@doi [\mnras] {10.1093/mnras/stw1059}, \href
  {http://adsabs.harvard.edu/abs/2016MNRAS.460.1645A} {460, 1645}

\bibitem[\protect\citeauthoryear{{Adams}, {Kochanek}, {Gerke}, {Stanek}  \&
  {Dai}}{{Adams} et~al.}{2017}]{Adams17}
{Adams} S.~M.,  {Kochanek} C.~S.,  {Gerke} J.~R.,  {Stanek} K.~Z.,   {Dai} X.,
  2017, \mn@doi [\mnras] {10.1093/mnras/stx816}, \href
  {http://adsabs.harvard.edu/abs/2017MNRAS.468.4968A} {468, 4968}

\bibitem[\protect\citeauthoryear{{Adams} et~al.,}{{Adams}
  et~al.}{2018}]{Adams2018}
{Adams} S.~M.,  et~al., 2018, \mn@doi [\pasp] {10.1088/1538-3873/aaa356}, \href
  {http://adsabs.harvard.edu/abs/2018PASP..130c4202A} {130, 034202}

\bibitem[\protect\citeauthoryear{{Astropy Collaboration} et~al.,}{{Astropy
  Collaboration} et~al.}{2013}]{astropy:2013}
{Astropy Collaboration} et~al., 2013, \mn@doi [\aap]
  {10.1051/0004-6361/201322068}, \href
  {https://ui.adsabs.harvard.edu/abs/2013A&A...558A..33A} {558, A33}

\bibitem[\protect\citeauthoryear{{Astropy Collaboration} et~al.,}{{Astropy
  Collaboration} et~al.}{2018}]{astropy:2018}
{Astropy Collaboration} et~al., 2018, \mn@doi [\aj] {10.3847/1538-3881/aabc4f},
  \href {https://ui.adsabs.harvard.edu/abs/2018AJ....156..123A} {156, 123}

\bibitem[\protect\citeauthoryear{{Bersier}, {Kochanek}, {Wagner}, {Adams}  \&
  {Dong}}{{Bersier} et~al.}{2015}]{Bersier15}
{Bersier} D.,  {Kochanek} C.~S.,  {Wagner} R.~M.,  {Adams} S.,   {Dong} S.,
  2015, The Astronomer's Telegram, \href
  {http://adsabs.harvard.edu/abs/2015ATel.7537....1B} {7537}

\bibitem[\protect\citeauthoryear{{Bildsten}, {Shen}, {Weinberg}  \&
  {Nelemans}}{{Bildsten} et~al.}{2007}]{Bildsten2007ApJ}
{Bildsten} L.,  {Shen} K.~J.,  {Weinberg} N.~N.,   {Nelemans} G.,  2007,
  \mn@doi [\apjl] {10.1086/519489}, \href
  {https://ui.adsabs.harvard.edu/abs/2007ApJ...662L..95B} {662, L95}

\bibitem[\protect\citeauthoryear{{Blagorodnova} et~al.,}{{Blagorodnova}
  et~al.}{2017}]{Blagorodnova17}
{Blagorodnova} N.,  et~al., 2017, \mn@doi [\apj] {10.3847/1538-4357/834/2/107},
  \href {http://adsabs.harvard.edu/abs/2017ApJ...834..107B} {834, 107}

\bibitem[\protect\citeauthoryear{{Bond}, {Bedin}, {Bonanos}, {Humphreys},
  {Monard}, {Prieto}  \& {Walter}}{{Bond} et~al.}{2009}]{Bond2009}
{Bond} H.~E.,  {Bedin} L.~R.,  {Bonanos} A.~Z.,  {Humphreys} R.~M.,  {Monard}
  L.~A.~G.~B.,  {Prieto} J.~L.,   {Walter} F.~M.,  2009, \mn@doi [\apjl]
  {10.1088/0004-637X/695/2/L154}, \href
  {http://adsabs.harvard.edu/abs/2009ApJ...695L.154B} {695, L154}

\bibitem[\protect\citeauthoryear{{Botticella} et~al.,}{{Botticella}
  et~al.}{2009}]{Botticella2009MNRAS}
{Botticella} M.~T.,  et~al., 2009, \mn@doi [\mnras]
  {10.1111/j.1365-2966.2009.15082.x}, \href
  {https://ui.adsabs.harvard.edu/abs/2009MNRAS.398.1041B} {398, 1041}

\bibitem[\protect\citeauthoryear{{Chambers} et~al.,}{{Chambers}
  et~al.}{2016}]{Chambers16}
{Chambers} K.~C.,  et~al., 2016, arXiv:1612.05560, \href
  {http://adsabs.harvard.edu/abs/2016arXiv161205560C} {}

\bibitem[\protect\citeauthoryear{{Chemin}, {Carignan}  \& {Foster}}{{Chemin}
  et~al.}{2009}]{Chemin2009ApJ}
{Chemin} L.,  {Carignan} C.,   {Foster} T.,  2009, \mn@doi [\apj]
  {10.1088/0004-637X/705/2/1395}, \href
  {https://ui.adsabs.harvard.edu/abs/2009ApJ...705.1395C} {705, 1395}

\bibitem[\protect\citeauthoryear{{Darwin}}{{Darwin}}{1879}]{Darwin1879RSPS}
{Darwin} G.~H.,  1879, Proceedings of the Royal Society of London Series I,
  \href {https://ui.adsabs.harvard.edu/abs/1879RSPS...29..168D} {29, 168}

\bibitem[\protect\citeauthoryear{{Dominik}, {Belczynski}, {Fryer}, {Holz},
  {Berti}, {Bulik}, {Mand el}  \& {O'Shaughnessy}}{{Dominik}
  et~al.}{2012}]{Dominik2012ApJ}
{Dominik} M.,  {Belczynski} K.,  {Fryer} C.,  {Holz} D.~E.,  {Berti} E.,
  {Bulik} T.,  {Mand el} I.,   {O'Shaughnessy} R.,  2012, \mn@doi [\apj]
  {10.1088/0004-637X/759/1/52}, \href
  {https://ui.adsabs.harvard.edu/abs/2012ApJ...759...52D} {759, 52}

\bibitem[\protect\citeauthoryear{{Dong}, {Kochanek}, {Adams}  \&
  {Prieto}}{{Dong} et~al.}{2015}]{Dong15}
{Dong} S.,  {Kochanek} C.~S.,  {Adams} S.,   {Prieto} J.-L.,  2015, The
  Astronomer's Telegram, \href
  {http://adsabs.harvard.edu/abs/2015ATel.7173....1D} {7173}

\bibitem[\protect\citeauthoryear{{Draine} \& {Lee}}{{Draine} \&
  {Lee}}{1984}]{Draine84}
{Draine} B.~T.,  {Lee} H.~M.,  1984, \mn@doi [\apj] {10.1086/162480}, \href
  {http://adsabs.harvard.edu/abs/1984ApJ...285...89D} {285, 89}

\bibitem[\protect\citeauthoryear{{Draine} et~al.,}{{Draine}
  et~al.}{2014}]{Draine2014}
{Draine} B.~T.,  et~al., 2014, \mn@doi [\apj] {10.1088/0004-637X/780/2/172},
  \href {http://adsabs.harvard.edu/abs/2014ApJ...780..172D} {780, 172}

\bibitem[\protect\citeauthoryear{{Elitzur} \& {Ivezi{\'c}}}{{Elitzur} \&
  {Ivezi{\'c}}}{2001}]{Elitzur01}
{Elitzur} M.,  {Ivezi{\'c}} {\v Z}.,  2001, \mn@doi [MNRAS]
  {10.1046/j.1365-8711.2001.04706.x}, \href
  {http://adsabs.harvard.edu/abs/2001MNRAS.327..403E} {327, 403}

\bibitem[\protect\citeauthoryear{{Fabrika} et~al.,}{{Fabrika}
  et~al.}{2015}]{ATel6985}
{Fabrika} S.,  et~al., 2015, The Astronomer's Telegram, \href
  {https://ui.adsabs.harvard.edu/abs/2015ATel.6985....1F} {6985, 1}

\bibitem[\protect\citeauthoryear{Fazio et~al.,}{Fazio et~al.}{2004}]{Fazio2004}
Fazio G.~G.,  et~al., 2004, \mn@doi [The Astrophysical Journal Supplement
  Series] {10.1086/422843}, 154, 10

\bibitem[\protect\citeauthoryear{{Freedman} \& {Madore}}{{Freedman} \&
  {Madore}}{1990}]{Freedman1990}
{Freedman} W.~L.,  {Madore} B.~F.,  1990, \mn@doi [\apj] {10.1086/169469},
  \href {http://adsabs.harvard.edu/abs/1990ApJ...365..186F} {365, 186}

\bibitem[\protect\citeauthoryear{Gehrz et~al.,}{Gehrz et~al.}{2007}]{Gehrz2007}
Gehrz R.~D.,  et~al., 2007, \mn@doi [Review of Scientific Instruments]
  {10.1063/1.2431313}, 78, 011302

\bibitem[\protect\citeauthoryear{{Geier} \& {Pessev}}{{Geier} \&
  {Pessev}}{2015}]{ATel8220}
{Geier} S.,  {Pessev} P.,  2015, The Astronomer's Telegram, \href
  {https://ui.adsabs.harvard.edu/abs/2015ATel.8220....1G} {8220, 1}

\bibitem[\protect\citeauthoryear{{Giallongo} et~al.,}{{Giallongo}
  et~al.}{2008}]{Giallongo08}
{Giallongo} E.,  et~al., 2008, \mn@doi [\aap] {10.1051/0004-6361:20078402},
  \href {http://adsabs.harvard.edu/abs/2008A%26A...482..349G} {482, 349}

\bibitem[\protect\citeauthoryear{{Glanz} \& {Perets}}{{Glanz} \&
  {Perets}}{2018}]{Glanz18}
{Glanz} H.,  {Perets} H.~B.,  2018, \mn@doi [\mnras] {10.1093/mnrasl/sly065},
  \href {http://adsabs.harvard.edu/abs/2018MNRAS.478L..12G} {478, L12}

\bibitem[\protect\citeauthoryear{{Gunn} \& {Stryker}}{{Gunn} \&
  {Stryker}}{1983}]{GunnStryker1983}
{Gunn} J.~E.,  {Stryker} L.~L.,  1983, \mn@doi [\apjs] {10.1086/190861}, \href
  {https://ui.adsabs.harvard.edu/abs/1983ApJS...52..121G} {52, 121}

\bibitem[\protect\citeauthoryear{{Harmanen}, {McCollum}, {Laine}, {Rottler}  \&
  {Bruhweiler}}{{Harmanen} et~al.}{2015}]{ATel7595}
{Harmanen} J.,  {McCollum} B.,  {Laine} S.,  {Rottler} L.,   {Bruhweiler}
  F.~C.,  2015, The Astronomer's Telegram, \href
  {https://ui.adsabs.harvard.edu/abs/2015ATel.7595....1H} {7595, 1}

\bibitem[\protect\citeauthoryear{{Hill}, {Green}  \& {Slagle}}{{Hill}
  et~al.}{2006}]{Hill06}
{Hill} J.~M.,  {Green} R.~F.,   {Slagle} J.~H.,  2006, in Society of
  Photo-Optical Instrumentation Engineers (SPIE) Conference Series. p. 62670Y,
  \mn@doi{10.1117/12.669832}

\bibitem[\protect\citeauthoryear{{Hodgkin} et~al.,}{{Hodgkin}
  et~al.}{2015}]{Hodgkin2015ATel6952}
{Hodgkin} S.~T.,  et~al., 2015, The Astronomer's Telegram, \href
  {https://ui.adsabs.harvard.edu/abs/2015ATel.6952....1H} {6952, 1}

\bibitem[\protect\citeauthoryear{{Howitt}, {Stevenson}, {Vigna-G{\'o}mez},
  {Justham}, {Ivanova}, {Woods}, {Neijssel}  \& {Mandel}}{{Howitt}
  et~al.}{2020}]{Howitt2020MNRAS}
{Howitt} G.,  {Stevenson} S.,  {Vigna-G{\'o}mez} A.~r.,  {Justham} S.,
  {Ivanova} N.,  {Woods} T.~E.,  {Neijssel} C.~J.,   {Mandel} I.,  2020,
  \mn@doi [\mnras] {10.1093/mnras/stz3542}, \href
  {https://ui.adsabs.harvard.edu/abs/2020MNRAS.492.3229H} {492, 3229}

\bibitem[\protect\citeauthoryear{{Iaconi}, {Maeda}, {Nozawa}, {De Marco}  \&
  {Reichardt}}{{Iaconi} et~al.}{2020}]{Iaconi2020arXiv}
{Iaconi} R.,  {Maeda} K.,  {Nozawa} T.,  {De Marco} O.,   {Reichardt} T.,
  2020, arXiv e-prints, \href
  {https://ui.adsabs.harvard.edu/abs/2020arXiv200306151I} {p. arXiv:2003.06151}

\bibitem[\protect\citeauthoryear{{Ivanova} et~al.,}{{Ivanova}
  et~al.}{2013a}]{Ivanova2013araa}
{Ivanova} N.,  et~al., 2013a, \mn@doi [\aapr] {10.1007/s00159-013-0059-2},
  \href {http://adsabs.harvard.edu/abs/2013A%26ARv..21...59I} {21, 59}

\bibitem[\protect\citeauthoryear{{Ivanova}, {Justham}, {Avendano Nandez}  \&
  {Lombardi}}{{Ivanova} et~al.}{2013b}]{Ivanova13}
{Ivanova} N.,  {Justham} S.,  {Avendano Nandez} J.~L.,   {Lombardi} J.~C.,
  2013b, \mn@doi [Science] {10.1126/science.1225540}, \href
  {http://adsabs.harvard.edu/abs/2013Sci...339..433I} {339, 433}

\bibitem[\protect\citeauthoryear{{Ivezic} \& {Elitzur}}{{Ivezic} \&
  {Elitzur}}{1997}]{Ivezic97}
{Ivezic} Z.,  {Elitzur} M.,  1997, MNRAS, \href
  {http://adsabs.harvard.edu/abs/1997MNRAS.287..799I} {287, 799}

\bibitem[\protect\citeauthoryear{{Ivezic}, {Nenkova}  \& {Elitzur}}{{Ivezic}
  et~al.}{1999}]{Ivezic99}
{Ivezic} Z.,  {Nenkova} M.,   {Elitzur} M.,  1999, astro-ph/9910475, \href
  {http://adsabs.harvard.edu/abs/1999astro.ph.10475I} {}

\bibitem[\protect\citeauthoryear{{Izzard}, {Hall}, {Tauris}  \&
  {Tout}}{{Izzard} et~al.}{2012}]{Izzard2012IAUS}
{Izzard} R.~G.,  {Hall} P.~D.,  {Tauris} T.~M.,   {Tout} C.~A.,  2012, in IAU
  Symposium. pp 95--102, \mn@doi{10.1017/S1743921312010769}

\bibitem[\protect\citeauthoryear{{Jacoby}, {Hunter}  \& {Christian}}{{Jacoby}
  et~al.}{1984}]{Jacoby1984}
{Jacoby} G.~H.,  {Hunter} D.~A.,   {Christian} C.~A.,  1984, \mn@doi [\apjs]
  {10.1086/190983}, \href
  {https://ui.adsabs.harvard.edu/abs/1984ApJS...56..257J} {56, 257}

\bibitem[\protect\citeauthoryear{{Jencson} et~al.,}{{Jencson}
  et~al.}{2019a}]{Jencson2019}
{Jencson} J.~E.,  et~al., 2019a, \mn@doi [\apjl] {10.3847/2041-8213/ab2c05},
  \href {https://ui.adsabs.harvard.edu/abs/2019ApJ...880L..20J} {880, L20}

\bibitem[\protect\citeauthoryear{{Jencson} et~al.,}{{Jencson}
  et~al.}{2019b}]{Jencson2019ApJ}
{Jencson} J.~E.,  et~al., 2019b, \mn@doi [\apj] {10.3847/1538-4357/ab4a01},
  \href {https://ui.adsabs.harvard.edu/abs/2019ApJ...886...40J} {886, 40}

\bibitem[\protect\citeauthoryear{{Kami{\'n}ski} \& {Tylenda}}{{Kami{\'n}ski} \&
  {Tylenda}}{2011}]{Kaminski2011AA}
{Kami{\'n}ski} T.,  {Tylenda} R.,  2011, \mn@doi [\aap]
  {10.1051/0004-6361/201015950}, \href
  {https://ui.adsabs.harvard.edu/abs/2011A&A...527A..75K} {527, A75}

\bibitem[\protect\citeauthoryear{{Kami{\'n}ski}, {Mason}, {Tylenda}  \&
  {Schmidt}}{{Kami{\'n}ski} et~al.}{2015}]{Kaminski2015AA}
{Kami{\'n}ski} T.,  {Mason} E.,  {Tylenda} R.,   {Schmidt} M.~R.,  2015,
  \mn@doi [\aap] {10.1051/0004-6361/201526212}, \href
  {https://ui.adsabs.harvard.edu/abs/2015A&A...580A..34K} {580, A34}

\bibitem[\protect\citeauthoryear{{Kami{\'n}ski} et~al.,}{{Kami{\'n}ski}
  et~al.}{2017}]{Kaminski2017AA}
{Kami{\'n}ski} T.,  et~al., 2017, \mn@doi [\aap] {10.1051/0004-6361/201629838},
  \href {https://ui.adsabs.harvard.edu/abs/2017A&A...599A..59K} {599, A59}

\bibitem[\protect\citeauthoryear{{Kasliwal}}{{Kasliwal}}{2012}]{Kasliwal2012PASA}
{Kasliwal} M.~M.,  2012, \mn@doi [\pasa] {10.1071/AS11061}, \href
  {https://ui.adsabs.harvard.edu/abs/2012PASA...29..482K} {29, 482}

\bibitem[\protect\citeauthoryear{{Kasliwal} et~al.,}{{Kasliwal}
  et~al.}{2012}]{Kasliwal2012ApJ}
{Kasliwal} M.~M.,  et~al., 2012, \mn@doi [\apj] {10.1088/0004-637X/755/2/161},
  \href {https://ui.adsabs.harvard.edu/abs/2012ApJ...755..161K} {755, 161}

\bibitem[\protect\citeauthoryear{{Kasliwal} et~al.,}{{Kasliwal}
  et~al.}{2017}]{Kasliwal2017ApJ}
{Kasliwal} M.~M.,  et~al., 2017, \mn@doi [\apj] {10.3847/1538-4357/aa6978},
  \href {https://ui.adsabs.harvard.edu/abs/2017ApJ...839...88K} {839, 88}

\bibitem[\protect\citeauthoryear{{Keenan}, {Garrison}  \& {Deutsch}}{{Keenan}
  et~al.}{1974}]{Keenan1974ApJS}
{Keenan} P.~C.,  {Garrison} R.~F.,   {Deutsch} A.~J.,  1974, \mn@doi [\apjs]
  {10.1086/190318}, \href
  {https://ui.adsabs.harvard.edu/abs/1974ApJS...28..271K} {28, 271}

\bibitem[\protect\citeauthoryear{{Klencki}, {Nelemans}, {Istrate}  \&
  {Chruslinska}}{{Klencki} et~al.}{2020}]{Klencki2020arXiv200611286K}
{Klencki} J.,  {Nelemans} G.,  {Istrate} A.~G.,   {Chruslinska} M.,  2020,
  arXiv e-prints, \href {https://ui.adsabs.harvard.edu/abs/2020arXiv200611286K}
  {p. arXiv:2006.11286}

\bibitem[\protect\citeauthoryear{{Kochanek}, {Adams}  \&
  {Belczynski}}{{Kochanek} et~al.}{2014}]{Kochanek14_mergers}
{Kochanek} C.~S.,  {Adams} S.~M.,   {Belczynski} K.,  2014, \mn@doi [\mnras]
  {10.1093/mnras/stu1226}, \href
  {http://adsabs.harvard.edu/abs/2014MNRAS.443.1319K} {443, 1319}

\bibitem[\protect\citeauthoryear{{Kulkarni}}{{Kulkarni}}{2013}]{Kulkarni13}
{Kulkarni} S.~R.,  2013, The Astronomer's Telegram, \href
  {http://adsabs.harvard.edu/abs/2013ATel.4807....1K} {4807}

\bibitem[\protect\citeauthoryear{{Kulkarni} et~al.,}{{Kulkarni}
  et~al.}{2007}]{Kulkarni07}
{Kulkarni} S.~R.,  et~al., 2007, \mn@doi [\nat] {10.1038/nature05822}, \href
  {http://adsabs.harvard.edu/abs/2007Natur.447..458K} {447, 458}

\bibitem[\protect\citeauthoryear{{Kurtenkov} et~al.,}{{Kurtenkov}
  et~al.}{2015a}]{Kurtenkov15}
{Kurtenkov} A.~A.,  et~al., 2015a, \mn@doi [\aap]
  {10.1051/0004-6361/201526564}, \href
  {http://adsabs.harvard.edu/abs/2015A%26A...578L..10K} {578, L10}

\bibitem[\protect\citeauthoryear{{Kurtenkov}, {Ovcharov}, {Nedialkov},
  {Kostov}, {Bachev}, {Dimitrova}, {Popov}  \& {Valcheva}}{{Kurtenkov}
  et~al.}{2015b}]{ATel6941}
{Kurtenkov} A.,  {Ovcharov} E.,  {Nedialkov} P.,  {Kostov} A.,  {Bachev} R.,
  {Dimitrova} R.~V.~M.,  {Popov} V.,   {Valcheva} A.,  2015b, The Astronomer's
  Telegram, \href {https://ui.adsabs.harvard.edu/abs/2015ATel.6941....1K}
  {6941, 1}

\bibitem[\protect\citeauthoryear{{Laor} \& {Draine}}{{Laor} \&
  {Draine}}{1993}]{Laor1993ApJ}
{Laor} A.,  {Draine} B.~T.,  1993, \mn@doi [\apj] {10.1086/172149}, \href
  {https://ui.adsabs.harvard.edu/abs/1993ApJ...402..441L} {402, 441}

\bibitem[\protect\citeauthoryear{{Law} et~al.,}{{Law} et~al.}{2009}]{Law09}
{Law} N.~M.,  et~al., 2009, \mn@doi [\pasp] {10.1086/648598}, \href
  {http://adsabs.harvard.edu/abs/2009PASP..121.1395L} {121, 1395}

\bibitem[\protect\citeauthoryear{{Law} et~al.,}{{Law} et~al.}{2010}]{Law10}
{Law} N.~M.,  et~al., 2010, in Ground-based and Airborne Instrumentation for
  Astronomy III. p. 77353M, \mn@doi{10.1117/12.857400}

\bibitem[\protect\citeauthoryear{{Lipunov} et~al.,}{{Lipunov}
  et~al.}{2017}]{Lipunov17}
{Lipunov} V.~M.,  et~al., 2017, arXiv:1704.08178, \href
  {http://adsabs.harvard.edu/abs/2017arXiv170408178L} {}

\bibitem[\protect\citeauthoryear{{Lodders}}{{Lodders}}{2003}]{Lodders2003ApJ}
{Lodders} K.,  2003, \mn@doi [\apj] {10.1086/375492}, \href
  {https://ui.adsabs.harvard.edu/abs/2003ApJ...591.1220L} {591, 1220}

\bibitem[\protect\citeauthoryear{{Lomb}}{{Lomb}}{1976}]{Lomb76}
{Lomb} N.~R.,  1976, \mn@doi [\apss] {10.1007/BF00648343}, \href
  {http://adsabs.harvard.edu/abs/1976Ap%26SS..39..447L} {39, 447}

\bibitem[\protect\citeauthoryear{{Lopez-Camara}, {Moreno Mendez}  \& {De
  Colle}}{{Lopez-Camara} et~al.}{2020}]{Lopez-Camara2020arXiv}
{Lopez-Camara} D.,  {Moreno Mendez} E.,   {De Colle} F.,  2020, arXiv e-prints,
  \href {https://ui.adsabs.harvard.edu/abs/2020arXiv200404158L} {p.
  arXiv:2004.04158}

\bibitem[\protect\citeauthoryear{{L{\"u}}, {Zhu}  \& {Podsiadlowski}}{{L{\"u}}
  et~al.}{2013}]{Lu2013ApJ}
{L{\"u}} G.,  {Zhu} C.,   {Podsiadlowski} P.,  2013, \mn@doi [\apj]
  {10.1088/0004-637X/768/2/193}, \href
  {https://ui.adsabs.harvard.edu/abs/2013ApJ...768..193L} {768, 193}

\bibitem[\protect\citeauthoryear{{MacLeod} \& {Loeb}}{{MacLeod} \&
  {Loeb}}{2020}]{MacLeodLoeb2020ApJ}
{MacLeod} M.,  {Loeb} A.,  2020, \mn@doi [\apj] {10.3847/1538-4357/ab89b6},
  \href {https://ui.adsabs.harvard.edu/abs/2020ApJ...895...29M} {895, 29}

\bibitem[\protect\citeauthoryear{{MacLeod}, {Macias}, {Ramirez-Ruiz},
  {Grindlay}, {Batta}  \& {Montes}}{{MacLeod} et~al.}{2017}]{MacLeod17}
{MacLeod} M.,  {Macias} P.,  {Ramirez-Ruiz} E.,  {Grindlay} J.,  {Batta} A.,
  {Montes} G.,  2017, \mn@doi [\apj] {10.3847/1538-4357/835/2/282}, \href
  {http://adsabs.harvard.edu/abs/2017ApJ...835..282M} {835, 282}

\bibitem[\protect\citeauthoryear{{MacLeod}, {Ostriker}  \& {Stone}}{{MacLeod}
  et~al.}{2018a}]{MacLeodOstriker2018ApJ}
{MacLeod} M.,  {Ostriker} E.~C.,   {Stone} J.~M.,  2018a, \mn@doi [\apj]
  {10.3847/1538-4357/aacf08}, \href
  {https://ui.adsabs.harvard.edu/abs/2018ApJ...863....5M} {863, 5}

\bibitem[\protect\citeauthoryear{{MacLeod}, {Ostriker}  \& {Stone}}{{MacLeod}
  et~al.}{2018b}]{MacLeodOstrikerStone2018ApJ}
{MacLeod} M.,  {Ostriker} E.~C.,   {Stone} J.~M.,  2018b, \mn@doi [\apj]
  {10.3847/1538-4357/aae9eb}, \href
  {https://ui.adsabs.harvard.edu/abs/2018ApJ...868..136M} {868, 136}

\bibitem[\protect\citeauthoryear{{Martini}, {Wagner}, {Tomaney}, {Rich}, {della
  Valle}  \& {Hauschildt}}{{Martini} et~al.}{1999}]{Martini99}
{Martini} P.,  {Wagner} R.~M.,  {Tomaney} A.,  {Rich} R.~M.,  {della Valle} M.,
    {Hauschildt} P.~H.,  1999, \mn@doi [\aj] {10.1086/300951}, \href
  {http://adsabs.harvard.edu/abs/1999AJ....118.1034M} {118, 1034}

\bibitem[\protect\citeauthoryear{{Masci} et~al.,}{{Masci}
  et~al.}{2017}]{Masci17}
{Masci} F.~J.,  et~al., 2017, \mn@doi [\pasp]
  {10.1088/1538-3873/129/971/014002}, \href
  {http://adsabs.harvard.edu/abs/2017PASP..129a4002M} {129, 014002}

\bibitem[\protect\citeauthoryear{{Mason}, {Diaz}, {Williams}, {Preston}  \&
  {Bensby}}{{Mason} et~al.}{2010}]{Mason2010AA}
{Mason} E.,  {Diaz} M.,  {Williams} R.~E.,  {Preston} G.,   {Bensby} T.,  2010,
  \mn@doi [\aap] {10.1051/0004-6361/200913610}, \href
  {https://ui.adsabs.harvard.edu/abs/2010A&A...516A.108M} {516, A108}

\bibitem[\protect\citeauthoryear{{Mathis}, {Rumpl}  \& {Nordsieck}}{{Mathis}
  et~al.}{1977}]{Mathis77}
{Mathis} J.~S.,  {Rumpl} W.,   {Nordsieck} K.~H.,  1977, \mn@doi [\apj]
  {10.1086/155591}, \href {http://adsabs.harvard.edu/abs/1977ApJ...217..425M}
  {217, 425}

\bibitem[\protect\citeauthoryear{{Mauerhan}, {Van Dyk}, {Johansson}, {Fox},
  {Filippenko}  \& {Graham}}{{Mauerhan} et~al.}{2018}]{Mauerhan2018MNRAS}
{Mauerhan} J.~C.,  {Van Dyk} S.~D.,  {Johansson} J.,  {Fox} O.~D.,
  {Filippenko} A.~V.,   {Graham} M.~L.,  2018, \mn@doi [\mnras]
  {10.1093/mnras/stx2500}, \href
  {https://ui.adsabs.harvard.edu/abs/2018MNRAS.473.3765M} {473, 3765}

\bibitem[\protect\citeauthoryear{{Metzger} \& {Pejcha}}{{Metzger} \&
  {Pejcha}}{2017}]{Metzger2017}
{Metzger} B.~D.,  {Pejcha} O.,  2017, \mn@doi [\mnras] {10.1093/mnras/stx1768},
  \href {http://adsabs.harvard.edu/abs/2017MNRAS.471.3200M} {471, 3200}

\bibitem[\protect\citeauthoryear{{Montalto}, {Seitz}, {Riffeser}, {Hopp}, {Lee}
   \& {Sch{\"o}nrich}}{{Montalto} et~al.}{2009}]{Montalto2009}
{Montalto} M.,  {Seitz} S.,  {Riffeser} A.,  {Hopp} U.,  {Lee} C.-H.,
  {Sch{\"o}nrich} R.,  2009, \mn@doi [\aap] {10.1051/0004-6361/200912179},
  \href {http://adsabs.harvard.edu/abs/2009A%26A...507..283M} {507, 283}

\bibitem[\protect\citeauthoryear{{Moreno M{\'e}ndez}, {L{\'o}pez-C{\'a}mara}
  \& {De Colle}}{{Moreno M{\'e}ndez} et~al.}{2017}]{MorenoMendez2017MNRAS}
{Moreno M{\'e}ndez} E.,  {L{\'o}pez-C{\'a}mara} D.,   {De Colle} F.,  2017,
  \mn@doi [\mnras] {10.1093/mnras/stx1385}, \href
  {https://ui.adsabs.harvard.edu/abs/2017MNRAS.470.2929M} {470, 2929}

\bibitem[\protect\citeauthoryear{{Mould} et~al.,}{{Mould}
  et~al.}{1990}]{Mould1990ApJ}
{Mould} J.,  et~al., 1990, \mn@doi [\apjl] {10.1086/185702}, \href
  {https://ui.adsabs.harvard.edu/abs/1990ApJ...353L..35M} {353, L35}

\bibitem[\protect\citeauthoryear{{Munari} et~al.,}{{Munari}
  et~al.}{2002}]{Munari2002AA}
{Munari} U.,  et~al., 2002, \mn@doi [\aap] {10.1051/0004-6361:20020715}, \href
  {https://ui.adsabs.harvard.edu/abs/2002A&A...389L..51M} {389, L51}

\bibitem[\protect\citeauthoryear{{Oke} et~al.,}{{Oke} et~al.}{1995}]{Oke95}
{Oke} J.~B.,  et~al., 1995, \mn@doi [\pasp] {10.1086/133562}, \href
  {http://adsabs.harvard.edu/abs/1995PASP..107..375O} {107, 375}

\bibitem[\protect\citeauthoryear{{Ossenkopf} \& {Henning}}{{Ossenkopf} \&
  {Henning}}{1994}]{Ossenkopf1994}
{Ossenkopf} V.,  {Henning} T.,  1994, \aap, \href
  {https://ui.adsabs.harvard.edu/abs/1994A&A...291..943O} {291, 943}

\bibitem[\protect\citeauthoryear{{Ovcharov}, {Kurtenkov}, {Valcheva}  \&
  {Nedialkov}}{{Ovcharov} et~al.}{2015}]{ATel6924}
{Ovcharov} E.,  {Kurtenkov} A.,  {Valcheva} A.,   {Nedialkov} P.,  2015, The
  Astronomer's Telegram, \href
  {https://ui.adsabs.harvard.edu/abs/2015ATel.6924....1O} {6924, 1}

\bibitem[\protect\citeauthoryear{{Paczynski}}{{Paczynski}}{1976}]{Paczynski1976IAUS}
{Paczynski} B.,  1976, in {Eggleton} P.,  {Mitton} S.,   {Whelan} J.,  eds,
  IAU Symposium Vol. 73, Structure and Evolution of Close Binary Systems. p.~75

\bibitem[\protect\citeauthoryear{{Pastorello} \& {Fraser}}{{Pastorello} \&
  {Fraser}}{2019}]{Pastorello2019NatAs}
{Pastorello} A.,  {Fraser} M.,  2019, \mn@doi [Nature Astronomy]
  {10.1038/s41550-019-0809-9}, \href
  {https://ui.adsabs.harvard.edu/abs/2019NatAs...3..676P} {3, 676}

\bibitem[\protect\citeauthoryear{{Pastorello} et~al.,}{{Pastorello}
  et~al.}{2019a}]{Pastorello2019b}
{Pastorello} A.,  et~al., 2019a, \mn@doi [\aap] {10.1051/0004-6361/201935511},
  \href {https://ui.adsabs.harvard.edu/abs/2019A%26A...625L...8P} {625, L8}

\bibitem[\protect\citeauthoryear{{Pastorello} et~al.,}{{Pastorello}
  et~al.}{2019b}]{Pastorello2019a}
{Pastorello} A.,  et~al., 2019b, \mn@doi [\aap] {10.1051/0004-6361/201935999},
  \href {https://ui.adsabs.harvard.edu/abs/2019A&A...630A..75P} {630, A75}

\bibitem[\protect\citeauthoryear{{Pejcha}}{{Pejcha}}{2014}]{Pejcha14}
{Pejcha} O.,  2014, \mn@doi [\apj] {10.1088/0004-637X/788/1/22}, \href
  {http://adsabs.harvard.edu/abs/2014ApJ...788...22P} {788, 22}

\bibitem[\protect\citeauthoryear{{Pejcha}, {Metzger}  \& {Tomida}}{{Pejcha}
  et~al.}{2016a}]{Pejcha16b}
{Pejcha} O.,  {Metzger} B.~D.,   {Tomida} K.,  2016a, \mn@doi [\mnras]
  {10.1093/mnras/stv2592}, \href
  {http://adsabs.harvard.edu/abs/2016MNRAS.455.4351P} {455, 4351}

\bibitem[\protect\citeauthoryear{{Pejcha}, {Metzger}  \& {Tomida}}{{Pejcha}
  et~al.}{2016b}]{Pejcha16}
{Pejcha} O.,  {Metzger} B.~D.,   {Tomida} K.,  2016b, \mn@doi [\mnras]
  {10.1093/mnras/stw1481}, \href
  {http://adsabs.harvard.edu/abs/2016MNRAS.461.2527P} {461, 2527}

\bibitem[\protect\citeauthoryear{{Pejcha}, {Metzger}, {Tyles}  \&
  {Tomida}}{{Pejcha} et~al.}{2017}]{Pejcha2017ApJ}
{Pejcha} O.,  {Metzger} B.~D.,  {Tyles} J.~G.,   {Tomida} K.,  2017, \mn@doi
  [\apj] {10.3847/1538-4357/aa95b9}, \href
  {https://ui.adsabs.harvard.edu/abs/2017ApJ...850...59P} {850, 59}

\bibitem[\protect\citeauthoryear{{Pessev}, {Geier}, {Kurtenkov}, {Nielsen}  \&
  {Tomov}}{{Pessev} et~al.}{2015a}]{ATel7272}
{Pessev} P.,  {Geier} S.,  {Kurtenkov} A.,  {Nielsen} L.~D.,   {Tomov} T.,
  2015a, The Astronomer's Telegram, \href
  {https://ui.adsabs.harvard.edu/abs/2015ATel.7272....1P} {7272, 1}

\bibitem[\protect\citeauthoryear{{Pessev}, {Geier}, {Kurtenkov}, {Nielsen},
  {Slumstrup}  \& {Tomov}}{{Pessev} et~al.}{2015b}]{ATel7572}
{Pessev} P.,  {Geier} S.,  {Kurtenkov} A.,  {Nielsen} L.~D.,  {Slumstrup} D.,
  {Tomov} T.,  2015b, The Astronomer's Telegram, \href
  {https://ui.adsabs.harvard.edu/abs/2015ATel.7572....1P} {7572, 1}

\bibitem[\protect\citeauthoryear{{Pessev}, {Geier}, {Stritzinger}, {Kurtenkov}
  \& {Tomov}}{{Pessev} et~al.}{2015c}]{ATel7624}
{Pessev} P.,  {Geier} S.,  {Stritzinger} M.,  {Kurtenkov} A.,   {Tomov} T.,
  2015c, The Astronomer's Telegram, \href
  {https://ui.adsabs.harvard.edu/abs/2015ATel.7624....1P} {7624, 1}

\bibitem[\protect\citeauthoryear{{Pessev}, {Geier}, {Stritzinger}, {Kurtenkov}
  \& {Tomov}}{{Pessev} et~al.}{2015d}]{ATel8059}
{Pessev} P.,  {Geier} S.,  {Stritzinger} M.,  {Kurtenkov} A.,   {Tomov} T.,
  2015d, The Astronomer's Telegram, \href
  {https://ui.adsabs.harvard.edu/abs/2015ATel.8059....1P} {8059, 1}

\bibitem[\protect\citeauthoryear{{Prieto}, {Sellgren}, {Thompson}  \&
  {Kochanek}}{{Prieto} et~al.}{2009}]{Prieto09}
{Prieto} J.~L.,  {Sellgren} K.,  {Thompson} T.~A.,   {Kochanek} C.~S.,  2009,
  \mn@doi [\apj] {10.1088/0004-637X/705/2/1425}, \href
  {http://adsabs.harvard.edu/abs/2009ApJ...705.1425P} {705, 1425}

\bibitem[\protect\citeauthoryear{{Rahmer}, {Smith}, {Velur}, {Hale}, {Law},
  {Bui}, {Petrie}  \& {Dekany}}{{Rahmer} et~al.}{2008}]{Rahmer08}
{Rahmer} G.,  {Smith} R.,  {Velur} V.,  {Hale} D.,  {Law} N.,  {Bui} K.,
  {Petrie} H.,   {Dekany} R.,  2008, in Ground-based and Airborne
  Instrumentation for Astronomy II. p. 70144Y, \mn@doi{10.1117/12.788086}

\bibitem[\protect\citeauthoryear{{Rasio}}{{Rasio}}{1995}]{Rasio1995ApJ}
{Rasio} F.~A.,  1995, \mn@doi [\apjl] {10.1086/187855}, \href
  {https://ui.adsabs.harvard.edu/abs/1995ApJ...444L..41R} {444, L41}

\bibitem[\protect\citeauthoryear{{Rau} et~al.,}{{Rau} et~al.}{2009}]{Rau09}
{Rau} A.,  et~al., 2009, \mn@doi [\pasp] {10.1086/605911}, \href
  {http://adsabs.harvard.edu/abs/2009PASP..121.1334R} {121, 1334}

\bibitem[\protect\citeauthoryear{{Reichardt}, {De Marco}, {Iaconi}, {Tout}  \&
  {Price}}{{Reichardt} et~al.}{2019}]{Reichardt2019MNRAS}
{Reichardt} T.~A.,  {De Marco} O.,  {Iaconi} R.,  {Tout} C.~A.,   {Price}
  D.~J.,  2019, \mn@doi [\mnras] {10.1093/mnras/sty3485}, \href
  {https://ui.adsabs.harvard.edu/abs/2019MNRAS.484..631R} {484, 631}

\bibitem[\protect\citeauthoryear{{Rich}, {Mould}, {Picard}, {Frogel}  \&
  {Davies}}{{Rich} et~al.}{1989}]{Rich89}
{Rich} R.~M.,  {Mould} J.,  {Picard} A.,  {Frogel} J.~A.,   {Davies} R.,  1989,
  \mn@doi [\apjl] {10.1086/185455}, \href
  {http://adsabs.harvard.edu/abs/1989ApJ...341L..51R} {341, L51}

\bibitem[\protect\citeauthoryear{{Scargle}}{{Scargle}}{1982}]{Scargle82}
{Scargle} J.~D.,  1982, \mn@doi [\apj] {10.1086/160554}, \href
  {http://adsabs.harvard.edu/abs/1982ApJ...263..835S} {263, 835}

\bibitem[\protect\citeauthoryear{{Schlafly} et~al.,}{{Schlafly}
  et~al.}{2012}]{Schlafly12}
{Schlafly} E.~F.,  et~al., 2012, \mn@doi [\apj] {10.1088/0004-637X/756/2/158},
  \href {http://adsabs.harvard.edu/abs/2012ApJ...756..158S} {756, 158}

\bibitem[\protect\citeauthoryear{{Schlegel}, {Finkbeiner}  \&
  {Davis}}{{Schlegel} et~al.}{1998}]{Schlegel98}
{Schlegel} D.~J.,  {Finkbeiner} D.~P.,   {Davis} M.,  1998, \mn@doi [ApJ]
  {10.1086/305772}, \href {http://adsabs.harvard.edu/abs/1998ApJ...500..525S}
  {500, 525}

\bibitem[\protect\citeauthoryear{{Seifert} et~al.,}{{Seifert}
  et~al.}{2003}]{Seifert03}
{Seifert} W.,  et~al., 2003, in {Iye} M.,  {Moorwood} A.~F.~M.,  eds,
  \procspie Vol. 4841, Instrument Design and Performance for Optical/Infrared
  Ground-based Telescopes. pp 962--973, \mn@doi{10.1117/12.459494}

\bibitem[\protect\citeauthoryear{{Shiber} \& {Soker}}{{Shiber} \&
  {Soker}}{2018}]{ShiberSoker2018MNRAS}
{Shiber} S.,  {Soker} N.,  2018, \mn@doi [\mnras] {10.1093/mnras/sty843}, \href
  {https://ui.adsabs.harvard.edu/abs/2018MNRAS.477.2584S} {477, 2584}

\bibitem[\protect\citeauthoryear{{Shiber}, {Iaconi}, {De Marco}  \&
  {Soker}}{{Shiber} et~al.}{2019}]{Shiber2019MNRAS}
{Shiber} S.,  {Iaconi} R.,  {De Marco} O.,   {Soker} N.,  2019, \mn@doi
  [\mnras] {10.1093/mnras/stz2013}, \href
  {https://ui.adsabs.harvard.edu/abs/2019MNRAS.488.5615S} {488, 5615}

\bibitem[\protect\citeauthoryear{{Shumkov} et~al.,}{{Shumkov}
  et~al.}{2015a}]{Shumkov15}
{Shumkov} V.,  et~al., 2015a, The Astronomer's Telegram, \href
  {http://adsabs.harvard.edu/abs/2015ATel.6911....1S} {6911}

\bibitem[\protect\citeauthoryear{{Shumkov} et~al.,}{{Shumkov}
  et~al.}{2015b}]{ATel6951}
{Shumkov} V.,  et~al., 2015b, The Astronomer's Telegram, \href
  {https://ui.adsabs.harvard.edu/abs/2015ATel.6951....1S} {6951, 1}

\bibitem[\protect\citeauthoryear{{Smith}, {Li}, {Silverman}, {Ganeshalingam}
  \& {Filippenko}}{{Smith} et~al.}{2011}]{Smith2011MNRAS}
{Smith} N.,  {Li} W.,  {Silverman} J.~M.,  {Ganeshalingam} M.,   {Filippenko}
  A.~V.,  2011, \mn@doi [\mnras] {10.1111/j.1365-2966.2011.18763.x}, \href
  {https://ui.adsabs.harvard.edu/abs/2011MNRAS.415..773S} {415, 773}

\bibitem[\protect\citeauthoryear{{Smith} et~al.,}{{Smith}
  et~al.}{2016}]{Smith16}
{Smith} N.,  et~al., 2016, \mn@doi [\mnras] {10.1093/mnras/stw219}, \href
  {http://adsabs.harvard.edu/abs/2016MNRAS.458..950S} {458, 950}

\bibitem[\protect\citeauthoryear{{Srivastava}, {Ashok}, {Banerjee}  \&
  {Venkataraman}}{{Srivastava} et~al.}{2015}]{ATel7236}
{Srivastava} M.,  {Ashok} N.~M.,  {Banerjee} D.~P.~K.,   {Venkataraman} V.,
  2015, The Astronomer's Telegram, \href
  {https://ui.adsabs.harvard.edu/abs/2015ATel.7236....1S} {7236, 1}

\bibitem[\protect\citeauthoryear{{Steele}, {Williams}, {Darnley}, {Bode},
  {Barnsley}, {Smith}  \& {Jermak}}{{Steele} et~al.}{2015}]{ATel7555}
{Steele} I.~A.,  {Williams} S.~C.,  {Darnley} M.~J.,  {Bode} M.~F.,  {Barnsley}
  R.~M.,  {Smith} R.~J.,   {Jermak} H.~E.,  2015, The Astronomer's Telegram,
  \href {https://ui.adsabs.harvard.edu/abs/2015ATel.7555....1S} {7555, 1}

\bibitem[\protect\citeauthoryear{{Thompson}, {Prieto}, {Stanek}, {Kistler},
  {Beacom}  \& {Kochanek}}{{Thompson} et~al.}{2009}]{Thompson09}
{Thompson} T.~A.,  {Prieto} J.~L.,  {Stanek} K.~Z.,  {Kistler} M.~D.,  {Beacom}
  J.~F.,   {Kochanek} C.~S.,  2009, \mn@doi [ApJ]
  {10.1088/0004-637X/705/2/1364}, \href
  {http://adsabs.harvard.edu/abs/2009ApJ...705.1364T} {705, 1364}

\bibitem[\protect\citeauthoryear{{Tylenda}}{{Tylenda}}{2005}]{Tylenda05a}
{Tylenda} R.,  2005, \mn@doi [\aap] {10.1051/0004-6361:20052800}, \href
  {http://adsabs.harvard.edu/abs/2005A\%26A...436.1009T} {436, 1009}

\bibitem[\protect\citeauthoryear{{Tylenda} \& {Kami{\'n}ski}}{{Tylenda} \&
  {Kami{\'n}ski}}{2016}]{Tylenda2016}
{Tylenda} R.,  {Kami{\'n}ski} T.,  2016, \mn@doi [\aap]
  {10.1051/0004-6361/201527700}, \href
  {http://adsabs.harvard.edu/abs/2016A%26A...592A.134T} {592, A134}

\bibitem[\protect\citeauthoryear{{Tylenda}, {Crause}, {G{\'o}rny}  \&
  {Schmidt}}{{Tylenda} et~al.}{2005}]{Tylenda05b}
{Tylenda} R.,  {Crause} L.~A.,  {G{\'o}rny} S.~K.,   {Schmidt} M.~R.,  2005,
  \mn@doi [\aap] {10.1051/0004-6361:20041581}, \href
  {http://adsabs.harvard.edu/abs/2005A\%26A...439..651T} {439, 651}

\bibitem[\protect\citeauthoryear{{Tylenda} et~al.,}{{Tylenda}
  et~al.}{2011}]{Tylenda11}
{Tylenda} R.,  et~al., 2011, \mn@doi [\aap] {10.1051/0004-6361/201016221},
  \href {http://adsabs.harvard.edu/abs/2011A%26A...528A.114T} {528, A114}

\bibitem[\protect\citeauthoryear{{Tylenda} et~al.,}{{Tylenda}
  et~al.}{2013}]{Tylenda13}
{Tylenda} R.,  et~al., 2013, \mn@doi [\aap] {10.1051/0004-6361/201321647},
  \href {http://adsabs.harvard.edu/abs/2013A%26A...555A..16T} {555, A16}

\bibitem[\protect\citeauthoryear{{Vigna-G{\'o}mez}, {MacLeod}, {Neijssel},
  {Broekgaarden}, {Justham}, {Howitt}, {de Mink}  \& {Mand
  el}}{{Vigna-G{\'o}mez} et~al.}{2020}]{VignaGomez2020arXiv}
{Vigna-G{\'o}mez} A.,  {MacLeod} M.,  {Neijssel} C.~J.,  {Broekgaarden} F.~S.,
  {Justham} S.,  {Howitt} G.,  {de Mink} S.~E.,   {Mand el} I.,  2020, arXiv
  e-prints, \href {https://ui.adsabs.harvard.edu/abs/2020arXiv200109829V} {p.
  arXiv:2001.09829}

\bibitem[\protect\citeauthoryear{{Wagner}, {Starrfield}, {Wilber}, {Kochanek},
  {Dong}, {Prieto}  \& {Adams}}{{Wagner} et~al.}{2015}]{Wagner15}
{Wagner} R.~M.,  {Starrfield} S.~G.,  {Wilber} A.,  {Kochanek} C.~S.,  {Dong}
  S.,  {Prieto} J.-L.,   {Adams} S.,  2015, The Astronomer's Telegram, \href
  {http://adsabs.harvard.edu/abs/2015ATel.7208....1W} {7208}

\bibitem[\protect\citeauthoryear{Werner et~al.,}{Werner
  et~al.}{2004}]{Werner2004}
Werner M.~W.,  et~al., 2004, \mn@doi [The Astrophysical Journal Supplement
  Series] {10.1086/422992}, 154, 1

\bibitem[\protect\citeauthoryear{{Williams}, {Darnley}, {Bode}  \&
  {Steele}}{{Williams} et~al.}{2015}]{Williams2015}
{Williams} S.~C.,  {Darnley} M.~J.,  {Bode} M.~F.,   {Steele} I.~A.,  2015,
  \mn@doi [\apjl] {10.1088/2041-8205/805/2/L18}, \href
  {http://adsabs.harvard.edu/abs/2015ApJ...805L..18W} {805, L18}

\bibitem[\protect\citeauthoryear{{Wizinowich} et~al.,}{{Wizinowich}
  et~al.}{2006}]{Wizinowich2006PASP}
{Wizinowich} P.~L.,  et~al., 2006, \mn@doi [\pasp] {10.1086/499290}, \href
  {https://ui.adsabs.harvard.edu/abs/2006PASP..118..297W} {118, 297}

\bibitem[\protect\citeauthoryear{{Wozniak}}{{Wozniak}}{2000}]{Wozniak2000AcA}
{Wozniak} P.~R.,  2000, \actaa, \href
  {https://ui.adsabs.harvard.edu/abs/2000AcA....50..421W} {50, 421}

\bibitem[\protect\citeauthoryear{{Yaron} \& {Gal-Yam}}{{Yaron} \&
  {Gal-Yam}}{2012}]{YaronGal-Yam2012}
{Yaron} O.,  {Gal-Yam} A.,  2012, \mn@doi [\pasp] {10.1086/666656}, \href
  {http://adsabs.harvard.edu/abs/2012PASP..124..668Y} {124, 668}

\makeatother
\end{thebibliography}
\bibliographystyle{mnras}

\bsp	
\label{lastpage}
\end{document}